\documentclass[11pt]{article}
\pdfoutput=1
\usepackage[utf8]{inputenc} % allow utf-8 input
\usepackage{booktabs}       % professional-quality tables
\usepackage{amstext} % for \text macro
\usepackage{array}   % for \newcolumntype macro
\newcolumntype{R}{>{$}r<{$}} % math-mode version of "l" column type
\usepackage{amssymb, mathtools, xcolor,makecell,enumerate}
\usepackage{amsmath, amsthm, accents}

\DeclareMathOperator*{\arglexmin}{arg\,lex\,min}
\usepackage[english]{babel}
\usepackage{orcidlink}
\usepackage{caption, float}
\usepackage{pdflscape}

\usepackage[noend, ruled, linesnumbered, noline, longend]{algorithm2e}
\SetNlSty{bfseries}{\color{black}}{}
\usepackage{enumerate}
\usepackage{multirow}
\usepackage{scrlayer}
\usepackage{tcolorbox}
\usepackage{makecell}
\usepackage{xcolor}
\usepackage{tikz}
\usepackage{subcaption}
\usepackage{hyperref}
\usepackage{cleveref}

\usepackage{easyReview}
\usetikzlibrary{calc}
\usetikzlibrary{positioning}
\usepackage{url}
\pgfdeclarelayer{bg} 
\pgfsetlayers{bg,main}

\usepackage{natbib}

\newcommand{\N}{\mathbb{N}}

\newcommand{\NQP}{\mathcal{NQP}}

\newcommand{\dom}{\preceq_{D}}

\newcommand{\lexsmaller}{\prec_{lex}}
\newcommand{\lexsmallereq}{\preceq_{lex}}

\newcommand{\incoming}[1]{\delta^{-}(#1)}
\newcommand{\outgoing}[1]{\delta^{+}(#1)}
\newcommand{\bigoh}[1]{\mathcal{O}\left(#1\right)}

\newcommand{\Null}{\mathtt{NULL}}

\newcommand{\mcs}[1]{T_{G(#1)}^*}

\newcommand{\ts}{\textsuperscript}
\newcommand{\transitionGraph}{\mathcal{G}}
\newcommand{\transitionNodes}{\mathcal{V}}
\newcommand{\transitionArcs}{\mathcal{A}}
\newcommand{\refAlgo}[1]{\hyperref[algo:#1]{#1}}

\SetCommentSty{mycommfont}

\newcommand{\stateSpace}{\mathbb{R}^d_{\geq0}}

\definecolor{mintgreen}{rgb}{0.0, 0.47, 0.44}
\definecolor{alizarin}{rgb}{0.82, 0.1, 0.26}
\definecolor{applegreen}{rgb}{0.55, 0.71, 0.0}

\newtheorem{theorem}[]{Theorem}[]

\newtheorem{lemma}[theorem]{Lemma}

\theoremstyle{definition}
\newtheorem{definition}[theorem]{Definition}
\newtheorem{example}[theorem]{Example}
\newtheorem{remark}{Remark}

\usetikzlibrary{positioning}
\usetikzlibrary{arrows}
\usetikzlibrary{calc}
\tikzset{multiArcLabelAbove/.cd,
	shifted path label/.style={pos=0.5,fill=white, draw=none,rectangle,#1,sloped}}

\tikzset{shifted path/.style args={from #1 to #2 by #3}{insert path={
			let \p1=($(#1.east)-(#1.center)$),
			\p2=($(#2.east)-(#2.center)$),\p3=($(#1.center)-(#2.center)$),
			\n1={veclen(\x1,\y1)},\n2={veclen(\x2,\y2)},\n3={atan2(\y3,\x3)} in
			(#1.{\n3+180+asin(#3/\n1)}) to 
			(#2.{\n3-asin(#3/\n2)})
}}}
\tikzset{labeled shifted path/.style args={from #1 to #2 by #3 label #4 place #5}{insert path={
			let \p1=($(#1.east)-(#1.center)$),
			\p2=($(#2.east)-(#2.center)$),\p3=($(#1.center)-(#2.center)$),
			\n1={veclen(\x1,\y1)},\n2={veclen(\x2,\y2)},\n3={atan2(\y3,\x3)} in
			(#1.{\n3+180+asin(#3/\n1)}) to node[multiArcLabelAbove/shifted path label=#5]{\scriptsize{#4}}
			(#2.{\n3-asin(#3/\n2)})
}}} 

\hypersetup{
	pdftitle={Multiobjective Spanning Tree},
	pdfsubject={},
	pdfauthor={Pedro Maristany de las Casas},
	pdfkeywords={Multiobjective Spanning Tree},
}
        \title{New Dynamic Programming Algorithm for the Multiobjective Minimum Spanning Tree Problem}
    
        \author{Pedro Maristany de las Casas \and Antonio Sedeño-Noda \and Ralf Borndörfer}
        \date{May 01, 2023}
\begin{document}

     \maketitle
        \begin{abstract}
            The \emph{Multiobjective Minimum Spanning Tree} (MO-MST) problem is a variant of the Minimum Spanning Tree problem, in which the costs associated with every edge of the input graph are vectors. In this paper, we design a new dynamic programming MO-MST algorithm. Dynamic programming for a MO-MST instance leads to the definition of an instance of the One-to-One Multiobjective Shortest Path (MOSP) problem and both instances have equivalent solution sets. The arising MOSP instance is defined on a so called transition graph. We study the original size of this graph in detail and reduce its size using cost dependent arc pruning criteria. To solve the MOSP instance on the reduced transition graph, we design the Implicit Graph Multiobjective Dijkstra Algorithm (IG-MDA), exploiting recent improvements on MOSP algorithms from the literature. All in all, the new IG-MDA outperforms the current state of the art on a big set of instances from the literature. Our code and results are publicly available.
        \end{abstract}
    
    \section{Introduction}
    
	We consider the \emph{Multiobjective Minimum Spanning Tree} (MO-MST) problem. An instance of the problem is a tuple $(G, c, s)$ consisting of an undirected connected graph $G = (V,E)$, a $d$-dimensional edge cost functions $c: E \to \stateSpace{}$, and a node $s \in V$ that is assumed, w.l.o.g, to be the root of every spanning tree in $G$. Trees in $G$ are connected acyclic subgraphs. Moreover, we characterize any tree $t$ in $G$ by its edges, i.e., $t \subseteq E$. Then, the costs of $t$ are defined as $c(t) := \sum_{e \in t} c(e) \in \stateSpace{}$ and we refer to the set of nodes spanned by $t$ by $V(t) \subseteq V$. We use the notions \emph{efficiency} and \emph{non-dominance} to refer to \emph{optimal} trees.
	
	\begin{definition}
		\label{def:treeDominance}
		Consider a MO-MST instance $(G,c,s)$. Let $t,\ t'$ be trees in $G$. $t$ \emph{dominates} $t'$ if $V(t) = V(t')$, $c(t) \leq c(t')$, and $c(t) \neq c(t')$.
		Moreover, $t$ is an \emph{efficient} tree, if it is not dominated by any other tree. The cost vector of an efficient spanning tree of $V$ is called an \emph{nondominated cost vector}.
	\end{definition}

    In this paper we are interested in computing \emph{minimum complete sets of efficient trees}: a subset of the efficient spanning trees of a given graph $G$ w.r.t. an edge cost function $c$ s.t. for every nondominated cost vector, there is exactly one efficient tree in the subset. We use the $*$-operator to denote minimum complete sets of a given set of edges or trees.

    \begin{definition}
	\label{def:starOperator}
        For a discrete set $Y \subset \stateSpace$, $Y^* \subseteq Y$ is the unique set of nondominated vectors in $Y$.
        Consider an undirected graph $G = (V,E)$ and a $d$-dimensional edge cost function $c$. Let $X$ be a set of edges or trees in $G$ and define $c(X) := \{ c(x) \ | \ x \in X\} \subset \stateSpace{}$. Then, $c(X)^*$ is well defined and we define $X^* \subseteq X$ to be a subset of $X$ with minimum cardinality s.t. $c(X^*) = c(X)^*$.
	\end{definition}
    
    Formally we can now define the MO-MST as follows.

	\begin{definition}[Multiobjective Minimum Spanning Tree Problem]\label{def:momst}
		Consider an undirected graph $G = (V,E)$, a root node $s \in V$, and $d$-dimensional edge cost functions $c: E \to \stateSpace{}$. Let $T$ be the set of all spanning trees of $G$.
		The \emph{Multiobjective Minimum Spanning Tree} (MO-MST) problem is to find a minimum complete set $T^* \subseteq T$ of efficient trees according to \Cref{def:starOperator}. We refer to the pair $(G,c,s)$ as a \emph{$d$-dimensional MO-MST instance}.
	\end{definition}

	On a complete graph with $n \in \N$ labeled nodes there are $n^{n-2}$ spanning trees \citep{Cayley2009}. \citet{Hamacher1994} constructed bidimensional MO-MST instances on complete graphs in which every spanning tree has a nondominated cost vector, proving the problem's intractability. In their paper, they also include a result that proves the NP-hardness of the MO-MST problem.
	
	\subsection{Literature Review and Outline}

    The MO-MST problem is well studied in the literature. For good introduction to the topic with multiple insights on different solution approaches, we refer to \citep[Section 9.2.]{Ehrgott2005}. The recent survey by \citet{Fernandes2020} benchmarks different algorithms on a huge set of instances providing good insights on what techniques work best in practice nowadays.

    For the classical Minimum Spanning Tree problem, greedy approaches like the well known algorithms by \citet{Kruskal56} or by \citet{Prim57} are very efficient in theory and in practice. However, extrapolating these techniques to a multiobjective scenario like ours does not deliver good algorithmic performance. A good explanation is given in \citep[Section 9.2]{Ehrgott2005}. In a nutshell, the main reason is that with higher dimensional edge costs, cost-minimal edge to be included in every iteration are not well defined.
    
    This observation opens up the search for new algorithmic approaches to solve minimum spanning tree problems in multiobjective scenarios. Interestingly, this search leads to different answers depending on the problem's dimension. In the biobjective case two phase algorithms \citep{Hamacher1994, Ramos98, Steiner08, Sourd08, Amorosi22} have been proven to be most efficient. These algorithms compute the supported efficient spanning trees in the first phase and the unsupported efficient solutions in the second. Supported solutions are obtained solving multiple instances of the single criteria Minimum Spanning Tree problem via scalarization of the objective functions. For the computation of unsupported solutions, different techniques are used. In their state of the art algorithm \citet{Sourd08} split the second phase into two subphases. First they find unsupported solutions performing neighborhood search starting from the already found efficient trees. Finally the missing efficient trees are found solving multiple MIPs that use the costs of the existing efficient trees as bounds. The reason for the superb performance of this algorithm is that in almost every practical instance, all efficient trees are already computed after the neighborhood search. 

    For edge costs with more than two dimensions, efficient two phase approaches are not known as discussed for example in \citep{Ehrgott2000}. This once again opens up the search for efficient algorithms leading to dynamic programming algorithms. In this context, two publications by Di Puglia Pugliese, Guerreiro, and Santos stand out. They first published a dynamic programming algorithm for the MO-MST problem in \citep{DiPugliaPugliese2014} and afterwards they notably improved their approach in \citep{Santos18} designing the \emph{BN algorithm}. In it they include an elegant way of ensuring that every dynamically built tree is only considered once. By doing so, they solve an arising \emph{symmetry} issue in their original algorithm. 
    %For example, in a complete graph with three nodes $V = \{s, v, w\}$, the trees $t = ([s,v], [s,w])$ and $t' = ([s,w], [s,v])$ are equivalent and do not need to be considered twice. In the extensive benchmarks conducted in the survey by \citet{Fernandes2020} the BN algorithm turns out to be the most efficient exact algorithm for all kinds of MO-MST instances with more than two dimensional edge cost functions.

    % \alert{Much shorter!}Dynamic programming approaches are closely related to (multiobjective) shortest path algorithms. On a new directed graph, called the \emph{transition graph $\transitionGraph$}, every \emph{transition node} represents a subset of the nodes in the original input graph of the MO-MST instance $(G,c,s)$ (cf. \Cref{def:transitionGraph}). If we consider the node in the new graph representing the node set $U \subset V$, its outgoing arcs are copies of all edges in the cut $\delta(U) = \{[u,v] \in E \ | \ u \in U, v \notin U\}$ in $G$. Then, finding efficient paths from the transition node representing the node subset $\{s\}$ that only includes the root node to the node representing the whole node set $V$ of $G$ is equivalent to finding efficient spanning trees of $G$ (cf. \Cref{thm:treePathEquiv}).

    \paragraph{Outline} We start \Cref{sec:dynProgMoMST} embedding the MO-MST problem in a dynamic programming context. Then, similar to \citep[Section 3.1, Section 3.2]{Santos18}, we discuss how to transform a MO-MST instance into an instance of the One-to-One Multiobjective Shortest Path (MOSP) problem defined on a so called transition graph $\transitionGraph$. Both instances' solution sets are equivalent. In \Cref{sec:size} to \Cref{sec:chen} we discuss our first main contributions: how to reduce the size of $\transitionGraph$. The resulting One-to-One MOSP instance defined on the reduced transition graph is used in our second main contribution: the Implicit Graph Multiobjective Dijkstra Algorithm (IG-MDA) introduced in \Cref{sec:newAlgo}. It features multiple recent techniques to efficiently solve large scale MOSP instances \citep{Pulido14, Sedeno19, Maristany21, Casas22}. We also use these techniques in our implementation of the BN algorithm. It is described in \ref{sec:bn}. In \Cref{sec:experiments} we present the results of our computational experiments in which we benchmark the IG-MDA against the BN algorithm. Since on bidimensional MO-MST instances both algorithms are clearly outperformed by the two phase algorithm from \citet{Sourd08}, we use three and four dimensional instances in our experiments. We consider all such MO-MST instances from \citep{Santos18} and a big subset of the instances used in \citep{Fernandes2020}. Thereby, we achieve the last contribution in this paper: to the best of our knowledge the size of the solved instances is bigger than so far in the literature.
     
	\section{Dynamic Programming for MO-MST}
    \label{sec:dynProgMoMST}
	In dynamic programming, bellman conditions are recursive expressions that state how to derive efficient solutions for a problem at hand from efficient solutions of its subproblems. In our MO-MST scenario, the Bellman condition (cf. \Cref{thm:bellmanSubtrees}) states how to build efficient spanning subtrees of a subset $W \subseteq V$ by looking at the efficient spanning subtrees of all subsets $U \subset W$ with $|U| = |W|-1$. 

    Recall that any node set $U \subseteq V$ induces the \emph{cut} 
    \[ \delta(U) := \{[u,w] \in E \ | \ u \in U, \, w \in V\backslash U\} \]
    and we have the following basic result from graph theory.
 
	\begin{lemma}
		\label{lemma:growingTrees}
		Let $G = (V,E)$ be an undirected graph, $t$ a tree in $G$, and $[u,w] \in E$ an edge. Then, $t' = t \circ [u,w]$ is a tree if and only if $[u,w] \in \delta(V(t))$. If that is the case, we have $|V(t)| = |V(t')| - 1$.
	\end{lemma}

    We refer to the subgraph induced by a node set $U \subseteq V$ by $G(U)$. Additionally, the spanning trees of $G(U)$ are denoted by $T_{G(U)}$. Recall that the $*$-operator is defined in \Cref{def:starOperator}.
 
	\begin{theorem}[Bellman Condition for Efficient Subtrees]
		\label{thm:bellmanSubtrees}
		Let $(G = (V,E), c, s)$ be a $d$-dimensional MO-MST instance.

        \begin{description}
            \item[Base Case] We set $t_{G(\{s\})}$ to be the spanning tree of $G(\{s\})$, a subgraph containing just the node $s$ and no edges, and $c(t_{G(\{s\})}) = 0$. Since this is the only tree in $T_{G(\{s\})}$, we have $T_{G(\{s\})} = T_{G(\{s\})}^*$.
            \item[Recursion] Consider a subset $W \subseteq V$ of cardinality $k \in \{2, \dotsc, n\}$. A minimum complete set of efficient spanning trees of $G(W)$ is given by
            \begin{equation}
			    \label{eq:bellmanTrees}
			    T_{G(W)}^* = \Big\{ t' \circ [u,w] \ \Big| \ t' \in T_{G(U)}^*,\, U \subset W,\, |W| = |U|+1,\, \text{ and } [u,w] \in \delta(U) \Big\}^*.
		      \end{equation}
        \end{description}
	\end{theorem}

	\begin{proof}
 		%By \Cref{lemma:growingTrees} we know that $t \circ [u,v]$ can only be a spanning tree of $W$ if $[u,v] \in \delta(U)$ since $t$ is a (efficient) spanning tree of $U$. Moreover, the conditions in \eqref{eq:bellmanTrees} implicitly require that only subsets $U$ of $W$ are considered such that $W = U \cup \{v\}$ for some $v \notin U$. Thus, $|U| = |W|-1 $. \alert{Why do we need this?}
		
		Assume there exists an efficient spanning tree $t$ of $W$ and a subset $U \subset W$ s.t. $t = t'' \circ [u,w]$ for a dominated tree $t''$ of $U$. 
        %By \Cref{lemma:growingTrees} $[u,v]$ is a cut edge, i.e., $[u,v] \in \delta(U)$. 
        Consider an efficient spanning tree $t'$ of $U$ that dominates $t''$. $t' \circ [u,w]$ is also a spanning tree of $W$ (\Cref{lemma:growingTrees}) and
        \[
            c(t') \leq c(t'') \Leftrightarrow c(t' \circ [u,w]) \leq c(t'' \circ [u,w]) = c(t).
        \]
        Since $t'$ dominates $t''$ the first inequality is strict for at least one index causing $t$ to also be dominated. 
        By \Cref{lemma:growingTrees} every spanning tree of $W$ can be obtained from spanning trees of $U \subset W$ with $|U| + 1 = |W|$. Then, the correctness of the statement follows by induction over the cardinality of $|W|$ and using $T_{G(\{s\})} = T_{G(\{s\})}^*$ as the induction's base case.

        Note that the $*$-operator on the right hand side of \eqref{eq:bellmanTrees} is needed because the expansions of efficient trees of different subsets $U$ of $W$ can dominate each other.
	\end{proof}

    %Let $(G = (V,E), c)$ be a $d$-dimensional MO-MST instance. The new condition unveils how to design a so called \emph{transition graph} $\transitionGraph{} = (\transitionNodes, \transitionArcs{})$ in which a source node $s \in \transitionNodes{}$ represents the initial empty tree and the set of efficient $s$-$t$-paths associated with a target node $t \in \transitionNodes{}$ have a one to one correspondence with the spanning trees in a minimum complete set of efficient trees.

    In the remainder of this section, we discuss how to exploit the efficiency condition derived in \Cref{thm:bellmanSubtrees}. For a given MO-MST instance $(G = (V,E),c)$ we define a so called \emph{transition graph} $\transitionGraph = (\transitionNodes, \transitionArcs)$ where each \emph{transition node} in $\transitionNodes$ represents a subset of nodes in $G$. Our goal is to define $\transitionGraph$ in such a way that efficient paths from the transition node $\{s\}$ to the transition node that represents the whole node set $V$ correspond to efficient spanning trees of $G$.
    
    \begin{definition}[One-to-One Multiobjective Shortest Path Problem]
        \label{def:MOSP}
        Given a directed graph $\transitionNodes = (\transitionNodes, \transitionArcs)$, a cost function $\gamma : \transitionArcs \to \stateSpace{}$, a source node $\{s\} \in \transitionNodes$, and a target node $V \in \transitionNodes$, the One-to-One Multiobjective Shortest Path problem is to find a minimum complete set of efficient $\{s\}$-$V$-paths in $\transitionGraph$ w.r.t. $\gamma$. We refer to the tuple $(\transitionGraph, \gamma, \{s\}, V)$ as a problem's instance.
    \end{definition}

    Our exposition from \Cref{def:transitionGraph} to \Cref{thm:treePathEquiv} is similar to \citep[Section 3.1]{Santos18}.
    Note that for a discrete set $X$, the power set of $X$ is $2^X$. In the definition of $\transitionGraph$ we incur in a slight abuse of notation when denote a transition node $U$ and simultaneously mean the transition node in $\transitionNodes$ and the subset $U \subseteq V$ of nodes in the original graph $G$.

	\begin{definition}[Transition Graph]
        \label{def:transitionGraph}
		Consider a MO-MST instance $(G = (V, E),c,s)$. We define the \emph{transition graph of $G$} as the directed graph $\transitionGraph = (\transitionNodes, \transitionArcs)$ with node set $\transitionNodes := \{V' \cup \{s\} \  \Big| \  V' \in 2^{V\backslash \{s\}}\}$.
        Every such node is also called a \emph{transition node}.
		The outgoing arcs of node $U \in \transitionNodes$ are induced by the cut $\delta(U)$:
        \begin{equation}
            \label{eq:outgoingArcs}
            \outgoing{U} := \Big\{ (U,W) \ \Big| \ W = U \cup \{w\} 
            \text{ if } \exists [u,w] \in \delta(U) \Big\}.
        \end{equation}
        The set of arcs in $\transitionGraph$ is given by $\transitionArcs := \bigcup_{U \in \transitionNodes} \outgoing{U}$.
        For an arc $(U,W) \in \transitionArcs$ we denote the unique \emph{preimage edge} $[u,w] \in E$ that induces it by $(U,W)^{-1}$ and we refer to $(U,W)$ as a \emph{copy of $[u,w]$ in $\transitionGraph$}. 
        We define a $d$-dimensional arc cost function $\gamma: \transitionArcs \to \stateSpace{}$ setting $\gamma(a) = c(a^{-1})$.
	\end{definition}

    Note that for an arc $(U,W)$ as defined in \eqref{eq:outgoingArcs}, we have $U, W \in \transitionNodes$ and $U \cup \{w\} = W$ for some node $w \in V$, i.e., $W$ is an expansion of $U$ by just one node. Even though we defined the set $\transitionArcs$ using the nodes' outgoing arcs, it of course also allows us to access sets $\incoming{U}$ of incoming arcs for every transition node $U$. 
	The following remark is important for the understanding of the remainder of this section.

\begin{figure}
        \begin{tikzpicture}[every node/.style={draw=none, circle}, >=stealth']
                \node[] at (-2,0) (s) {s};
                \node[] at (2,0) (u) {u};
                \node[] at (0, 1) (w) {w};
                
                \draw[labeled shifted path=from s to u by 0pt label {$(3,3)$} place midway];
                \draw[labeled shifted path=from s to w by 0pt label {$(1,1)$} place midway];
                \draw[labeled shifted path=from u to w by 0pt label {$(2,1)$} place midway];

                \node[right=1cm of u] (1) {$\{s\}$};
                \node[above right= 1cm and 1cm of 1] (12) {$\{s,u\}$};
                \node[below right= 1cm and 1cm of 1] (13) {$\{s,v\}$};
                
                %\node[above right=of 12] (124) {$\{s,u,w\}$};
                \node[right= 3cm of 1] (123) {$\{s,u,v\} = V$};
                %\node[below right=of 13] (134) {$\{s,v,w\}$};
                
                %\node[right=2cm of 123] (1234) {$\{s,u,v,w\}=V$};
                
                \draw[->,labeled shifted path=from 1 to 12 by 0pt label {$(3,3)$} place midway];
                \draw[->,labeled shifted path=from 1 to 13 by 0pt label {$(1,1)$} place midway];
                
                %\draw[->,labeled shifted path=from 12 to 124 by -3pt label {$(2,1)$} place midway];
                
                \draw[->,labeled shifted path=from 12 to 123 by 3pt label {$(1,1)$} place above];
                \draw[->,labeled shifted path=from 12 to 123 by -3pt label {$(2,1)$} place below];
                
                \draw[->,labeled shifted path=from 13 to 123 by 3pt label {$(2,1)$} place above];
                \draw[->,labeled shifted path=from 13 to 123 by -3pt label {$(3,3)$} place below];
        \end{tikzpicture}
        \caption{Left: $2$d MO-MST instance $(G,c,s)$. Right: Corresponding One-to-One MOSP instance $(\transitionGraph, \gamma, \{s\}, V)$}\label{fig:example}
    \end{figure}
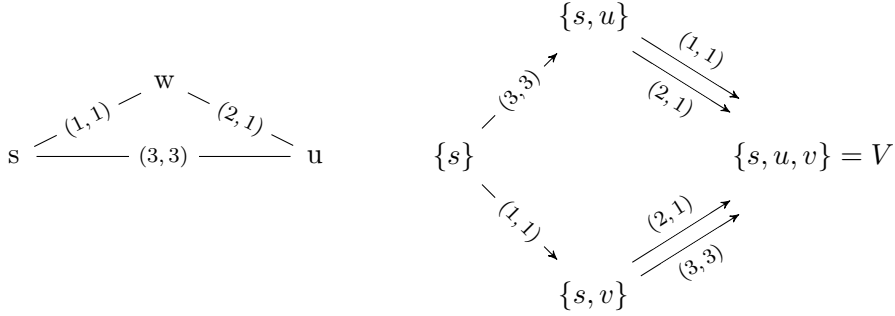

    \begin{remark}\label{rem:transitionProperties}
        The transition graph $\transitionGraph = (\transitionNodes{}, \transitionArcs{})$ as defined in \Cref{def:transitionGraph} is a layered acyclic multigraph.
        \begin{description}
            \item[Layered] For $k \in \{1, \cdots, n\}$ the $k$\ts{th} layer of $\transitionGraph$ consists of all nodes in $\transitionNodes{}$ that encode a subset of nodes in $V$ with cardinality $k$.
            \item[Acylic] From \eqref{eq:outgoingArcs} we immediately see that $\transitionArcs$ only contains arcs connecting neighboring layers and that the arcs always point towards the layer of greater cardinality.
            \item[Multigraph] $\transitionArcs$ is a multiset: multiple arcs connect the same pairs of nodes in $\transitionGraph$. Parallel arcs might have different costs depending on their preimage edge. Additionally, for two parallel arcs $a$, $b \in \transitionArcs$, there always holds $a^{-1} \neq b^{-1}$.
        \end{description}
        \Cref{fig:example} includes an example of a MO-MST instance and its corresponding One-to-One MOSP instance.
    \end{remark}

    The following definition formalizes the intuition of a path in $\transitionGraph$ \emph{representing} a tree in $G$.

    \begin{definition}[Path Representations of a tree]
        \label{def:pathRepresenations}
        Consider a MO-MST instance $(G,c,s)$ and the associated transition graph $\transitionGraph$. For a node $U \in \transitionNodes$ an $\{s\}$-$U$-path $p$ in $\transitionGraph$ is said to \emph{represent} or \emph{be a representation} of a spanning tree $t$ of $U$ in $G$ if $t = \{a^{-1} \ | \ a \in p \}$.
    \end{definition}

    The following lemma describes the mapping between trees in $G$ and paths in $\transitionGraph$ and their cost equivalence. Thereby it is important to note that a tree is represented by multiple paths but a path represents only one tree. The omitted proof corresponds to the proofs in \citep[Proposition 3.1., Proposition 3.2.]{Santos18}.

	\begin{lemma}
		\label{lem:treePathRepresentations}
		A tree $t$ in $G$ rooted at $s$ and spanning a node subset $U \subseteq V$ is represented in $\transitionGraph$ by at least one $\{s\}$-$U$-path. For any such path $p$ we have $c(t) = \gamma(p)$. Conversely, an $\{s\}$-$U$-path in $\transitionGraph$ for some $U \in \transitionNodes$ represents a tree $t$ in $G$ and $\gamma(p) = c(t)$.
	\end{lemma}

	We can now conclude that a given MO-MST instance can be solved computing the efficient paths in a One-to-One MOSP instance on the induced transition graph.
	
	\begin{theorem}
        \label{thm:treePathEquiv}
		Consider a subset $U \subseteq V$ that contains $s$. A minimum complete set of efficient $\{s\}$-$U$-paths in $\transitionGraph$ w.r.t. $\gamma$ corresponds to a minimum complete set of efficient spanning trees of $U$ in $G$ w.r.t. $c$.
	\end{theorem}

    \begin{proof}
        Let $P_U$ the set of $\{s\}$-$U$-paths in $\transitionGraph$. By \Cref{lem:treePathRepresentations} every such path represents a spanning tree of $U$ in $G$ and all spanning trees of $U$ are represented by at least one of these paths. Moreover, we have $\gamma(P_U) = c(T_{G(U)})$ and as a consequence, $\gamma(P_U)^* = c(T_{G(U)})^*$ which finishes the proof.
    \end{proof}
	
	\subsection{Size of the transition graph}
    \label{sec:size}
    In this section, we consider a $d$-dimensional MO-MST instance $(G=(V,E),\  c)$ in which $G$ is a complete graph with $n$ nodes. Additionally, we again assume that all spanning trees are rooted at a node $s \in V$ and build the equivalent One-to-One MOSP instance $(\transitionGraph, \gamma, \{s\}, V)$ as defined in \Cref{def:transitionGraph}.
    
    For a subset $U \subseteq V$ that contains $s$, let $t$ be a spanning tree of $G(U)$ with depth one.
    Then, there are $(|U|-1)!$ $\{s\}$-$U$-paths in $\transitionGraph$. 
    This number of representations of a tree in $\transitionGraph$ is an upper bound because for trees with depth greater than one, the ordering of the edges along the root-to-leaf paths in the tree have to be respected along the $\{s\}$-$U$-paths in $\transitionGraph$. If the spanning tree $t$ is a path, then the representation of $t$ in $\transitionGraph$ by a $\{s\}$-$U$-path is unique.

    In our setting, having fixed a node as the root node of all spanning trees, it is easy to see that for any $k \in \{1, \dots, n\}$ the $k$\ts{th} layer (cf. \Cref{rem:transitionProperties}) of nodes in $\transitionNodes$ is build out of $\binom{n-1}{k-1}$ nodes. Thus, we obtain
    \begin{equation}
        \label{eq:transitionNodesCount}
        |\transitionNodes| = \sum_{k=1}^{n} \binom{n-1}{k-1} = 2^{n-1}.
    \end{equation}
    
    The overall number of arcs between layer $k$ and layer $k+1$ for any $k \in \{1,\dotsc,n\}$ is 
    \begin{equation}
        \label{eq:transitionArcsPerLayer}
        \binom{n-1}{k-1} (n-k) k
    \end{equation}
    where the first multiplier is the number of transition nodes in the $k$\ts{th} layer, the second multiplier is the number of missing nodes from $G$ in any subset of nodes of $G$ with cardinality $k$, and the last multiplier accounts for the number of parallel arcs. Thus, all in all, the number of arcs in $\transitionGraph$ is 
    \begin{equation}
        \label{eq:transitionArcsCount}
        |\transitionArcs| = \sum_{k=1}^{n}\binom{n-1}{k-1} (n-k) k = 2^{n-3}(n-1)n
    \end{equation}

    Clearly, the size of the transition graph $\transitionGraph$ is an issue to address in the development of efficient dynamic programming algorithms for MO-MST problems. In the following sections, we discuss different techniques to reduce the size of $\transitionGraph$. In \Cref{sec:polyTimeRed}, we discuss a polynomial running time approach that acts on the input graph $G$ directly recognizing edges that can be deleted from the graph or edges that can be included in every efficient spanning tree. In \Cref{sec:domParallelArcs} and \Cref{sec:chen}, we discuss criteria to reduce the size of $\transitionArcs$. These criteria are cut-dependent and are applied during the construction of the sets $\outgoing{U}$ for $U \in \transitionNodes$.
    
    \subsection{Polynomial time graph reduction during preprocessing}
    \label{sec:polyTimeRed}

    The manipulation of the input graph $G$ described in this section is explained in \citep{Sourd08}. We refer the reader to the original paper to understand the details and correctness proofs.
    An edge $e \in E$ can be irrelevant for the MO-MST search, if $e$ is not contained in any efficient spanning tree of $G$ or if for every non dominated cost vector, there exists a spanning tree with this cost that contains $e$.
    
    \begin{definition}[Red and Blue Arcs. cf. \citep{Sourd08}, Section 3.2]
        Consider an instance $(G,c,s)$ of the MO-MST problem and let $T_G$ be the set of spanning trees of $G$ w.r.t. $c$.
        \begin{description}
            \item[Blue edge] An edge $e \in E$ is \emph{blue} if for every non-dominated cost vector in $c(T_G)^*$ there is an efficient tree $t \in T_G$ that contains $e$.
            \item[Red edge] An edge $e \in E$ is \emph{red} if it is not contained in any tree contained in a minimum complete set $T_G^*$ of efficient spanning trees of $G$.
        \end{description}
    \end{definition}

    Red edges can be deleted from $G$ without impacting the final set of efficient spanning trees. In case $e = [u,v]$ is determined to be a blue edge, $u$ and $v$ can be contracted to build a new node $w$ with $\delta(w) = (\delta(u) \cup \delta(v))\backslash \{[u,v]\}$. Once $w$ is built, $u$ and $v$ can be deleted from $G$.
    
    Interestingly, the conditions for an edge to be blue or red can be checked in polynomial time by running two Depth First Searches for every edge in $E$. Thus, as noted in \citet{Sourd08}, the preprocessing runs in $\bigoh{m^2}$.
    From now on, we only consider input graphs to our MO-MST instances that do not contain blue or red edges. By doing so, we assume that the described preprocessing phase is conducted before starting any actual MO-MST algorithm. 
    
    \subsection{Pruning parallel transition arcs}
    \label{sec:domParallelArcs}

    The red and the blue criteria for edges in $E$, reduce the size of the input graph $G$. In other words, a red or a blue edge in $G$ has no arc copies (cf. \Cref{def:transitionGraph}) in $\transitionGraph$. Using the condition that we derive in this section, we are only able to delete a subset of an edge's arc copies. In this sense, the condition from \Cref{lem:efficientParallelArcs} is more \emph{local} than the blue/red edge conditions.

    The condition is motivated by an almost trivial condition that holds for every One-to-One MOSP instance.

    \begin{lemma}
        Let $(\transitionGraph, \gamma, \{s\}, V)$ be a One-to-One MOSP instance. If $\transitionGraph$ contains two parallel arcs $a, \, a'$ s.t. $\gamma(a) \leq \gamma(a')$ and $\gamma(a) \neq \gamma(a')$ then $a'$ is not contained in any efficient $\{s\}$-$V$-path.
    \end{lemma}

    The condition becomes more powerful if we study the condition's meaning regarding the preimage edges of the involved arcs. 
    
    \begin{lemma}
        \label{lem:efficientParallelArcs}
        Let $U$ be a node set in $G$ containing $s$ and $[u, w]$, $[u', w]$ edges in $\delta(U)$ with $u, u' \in U$ and $w \in V \backslash U$. If $[u,w]$ dominates $[u', w]$, $[u', w]$ is not contained in any efficient spanning tree of $G$ that contains an efficient spanning tree of $U$ as a subtree.
    \end{lemma}
    
    \begin{proof}
    	Let $t_U$ be an efficient spanning tree of $U$ contained in an efficient spanning tree $t$ of $G$. If $t$ contains an edge $e = [u, w] \in \delta(U)$ with $u \in U$ and $w \notin U$ $e$ must not be dominated by any edge $e' = [u',w] \in \delta(U)$ because otherwise $t \circ e$ would be dominated by $t \circ e'$. Thus, $t \circ e$ would not be an efficient spanning tree of $U \cup \{w\}$ which contradicts \Cref{thm:bellmanSubtrees}.
    \end{proof}

    To explain the latter condition in terms of the transition graph, consider a transition node $U \in \transitionNodes$. We call every transition node $W \in \transitionNodes$ a \emph{successor node of $U$} if there exists a path from $U$ to $W$ in $\transitionGraph$. Since $\transitionGraph$ is acyclic (cf. \Cref{rem:transitionProperties}), this notion is well defined. Assume there exists a dominance relation between parallel outgoing arcs of $U$, i.e., between two arcs $a, a' \in \outgoing{U}$ whose head node is a transition node $W = U \cup \{w\}$ for some $w \in G$. Then, if $a'$ is dominated by $a$, no copies of the preimage edge $a'^{-1}$ need to be considered when building the sets $\outgoing{W'}$ for any transition node $W'$ that is a successor node of $U$. 

    \subsection{Pruning dominated outgoing arcs}
    \label{sec:chen}

    Our last condition to remove arcs from the transition graph is the most local one: if an arc $a \in \transitionArcs$ fulfills the condition in \Cref{lem:efficientCutExit}, we cannot gain further information concerning other arc copies from the preimage edge $a^{-1}$. Instead, the new condition only allows us to remove the arc $a$.

    % Recognizing red and blue edges gives us \emph{global} criteria to discard all arc copies of an edge in the transition graph. The condition from \Cref{lem:efficientParallelArcs} is more \emph{local} since it applies only to a subset of an edge's arc copies in the transition graph. More precisely, for an arc that meets this condition, we are able to look at its preimage edge and also remove other arc copies. In this section we again discuss a cut based arc removal condition.
    % It is the most \emph{local} condition considered in this paper because a deleted arc $(U,W)$, does not give us information about whether other arc copies of its preimage edge $(U,W)^{-1}$ can be also deleted.

    \begin{lemma}
        \label{lem:efficientCutExit}
        Consider a MO-MST instance $(G = (V,E), c)$ and assume w.l.o.g. that every spanning tree of $G$ is rooted at a fixed node $s \in V$. Let $t$ be an efficient spanning tree. For every edge $e \in t$ there is a cut $\delta(U)$ in $G$ and a minimum complete set $\delta(U)^*$ of efficient cut edges s.t. $e \in \delta(U)^*$.
    \end{lemma}

    \begin{proof}
        Fix an edge $e$ from $t$ and consider the two disjoint trees $t_U$ and $t_W$ obtained after removing $e$ from $t$. We assume that $t_U$ is the one containing the root node $s$ and set $U = V(t_U)$ to be its node set. Clearly, $e$ is a cut edge in $\delta(U)$ and any other cut edge $e'$ connects $t_U$ and $t_W$ hence building a new spanning tree $t'$ of $G$.
        If we suppose that a cut edge $e'$ in $\delta(U)$ dominates $e$ we get
        \[
            c(t) = c(t_U) + c(e) + c(t_W) \leq c(t_U) + c(e') + c(t_W) = c(t').
        \]
        and for at least one index $i \in \{1, \dots, d\}$, $c_i(t) < c_i(t')$ implying $t$'s dominance and thus contradicting the statement's assumption.
    \end{proof}

    A consequence of the last lemma is that when building the transition graph $\transitionGraph = (\transitionNodes, \transitionArcs)$ we do not need to include arc copies in $\outgoing{U}$ of edges that are dominated in the cut $\delta(U)$ of $G$. 
    Assume the edge $e \in \delta(U)$ is dominated but contained in an efficient spanning tree $t$ of $G$. \Cref{lem:efficientCutExit} guarantees that there is a node set $W \subset V$ for which $e$ is an efficient cut edge in $\delta(W)$. Then, $e$ is contained in a set of outgoing arcs $\outgoing{W}^*$ of the transition node $W \in \transitionNodes$. All in all, we obtain the following result.

    \begin{theorem}
	   Consider a MO-MST instance $(G=(V,E),c,s)$ without blue or red arcs in $G$. Any minimum complete set of efficient $\{s\}$-$V$-paths in the \emph{Pruned Transition Graph} $\transitionGraph^* = (\transitionNodes, \transitionArcs^*)$ with $\transitionArcs^* := \cup_{U \in \transitionNodes} \outgoing{U}^*$ corresponds to a minimum complete set of efficient $\{s\}$-$V$-paths in $\transitionGraph$.
    \end{theorem}
    
    \begin{remark}[Incoming Arcs]
        From now on, we refer to the sets of incoming arcs of a transition node $W \in \transitionNodes$ by $\incoming{W}^*$ but the notation can be misleading. We use it to emphasize that we only consider a transition graph with the set $\transitionArcs^*$ of arcs. However, even if the sets $\outgoing{U}^*$ do not contain dominated arcs for any $U \in \transitionNodes$, the set $\incoming{W}^* := \{(U,W) \ | \ (U,W) \in \transitionArcs^*\}$ can contain arcs that dominate each other.
    \end{remark}
    
    Note that \Cref{lem:efficientCutExit} is inspired by a pruning condition introduced in \citet[Section 3.2.]{Chen2007}. In fact, the pruned spanning subtrees therein and in our \Cref{lem:efficientCutExit} are the same. However, the authors in \citet{Chen2007} formulate the condition as a tree-dependent condition. For us it is important to formulate \Cref{lem:efficientCutExit} as a cut-dependent condition s.t. it needs only to be checked once for every transition node $U$, when building $\outgoing{U}^*$.
	
    \section{New Dynamic Programming Algorithm}
    \label{sec:newAlgo}
    
    In the last section, we described the translation from a given MO-MST instance $(G,c,s)$ with all trees rooted at a node $s \in V$ to the corresponding equivalent One-to-One MOSP instance $\mathcal{I}_{\text{SP}} = (\transitionGraph^*, \gamma, \{s\}, V)$. In this section, we explain how to use a slightly modified version of the \emph{Targeted Multiobjective Dijkstra Algorithm} (T-MDA) first published in \citep{Casas22} for One-to-One MOSP instances like $\mathcal{I}_{\text{SP}}$. The peculiarity of such instances is that in practice, the transition graphs actually needed to solve the instance are much smaller than the theoretical upper bounds derived in \eqref{eq:transitionNodesCount} and \eqref{eq:transitionArcsCount}.
    We are thus interested in a version of the T-MDA that does not store the transition graph explicitly. Instead, our new \emph{Implicit Graph Multiobjective Dijkstra Algorithm} (IG-MDA) creates the graph implicitly as needed during the algorithm's execution. We describe the algorithm using One-to-One MOSP nomenclature only. However, to emphasize the equivalence to the original MO-MST instances, we still use capital letters to denote the transition nodes in $\transitionGraph$, $\gamma$ for the arc cost function, and for $U \in \transitionNodes$, $T_{G(U)}^*$ to denote the minimum complete set of $\{s\}$-$U$-paths computed in $\transitionGraph$.
    
	Recall that we are interested in minimum complete sets of efficient paths (cf. \Cref{def:MOSP}). Thus, the IG-MDA only adds a new $\{s\}$-$U$-path $p$ to $T_{G(U)}^*$ if $p$ is not dominated by nor equivalent to any path therein. To achieve this algorithmically, we introduce the following $\dom{}$-operator.

    \begin{definition}
        Let $X \subset \stateSpace{}$ be a discrete set of cost vectors and $x' \in \stateSpace{}$. Then, $X \dom x'$ is true if and only it there is an $x \in X$ s.t. $x \leq x'$.
    \end{definition}

    Thus, $p$ is added to $T_{G(U)}^*$ only if $\gamma(T_{G(U)}^*) \dom \gamma(p)$ is not true. The lexicographic relation $\lexsmaller$ on $\stateSpace$ is defined as follows: for two vectors $x, \, y \in \stateSpace{}$, we write $x \lexsmaller y$ if $x_i < y_i$ for the first index $i \in \{1, \dotsc, d\}$ for which $x_i \neq y_i$. The relation $\lexsmallereq$ is defined analogously. 
    
    \Cref{algo:igmda} is the pseudocode of the IG-MDA. It is a label setting algorithm (cf. \citep{Maristany21}) for One-to-One Multiobjective Shortest Path problems. In our pseudocode, we assume the existence of a container $T^*$ that holds the lists $\mcs{U}$ for every $U \in \transitionNodes$. We discuss the details in the remainder to the section.
    
   	\begin{algorithm}
    	\small{
    		\SetKwInOut{Input}{Input} \SetKwInOut{Output}{Output}
    		\Input{$d$-dimensional One-to-One MOSP instance $\mathcal{I}_{\text{SP}} = (\transitionGraph^*, \gamma, \{s\}, V)$.}
    		\BlankLine
    		\Output{Minimal complete set $\mcs{V}$ of $\{s\}$-$V$-paths in $\transitionGraph$ w.r.t. $\gamma$.}
    		\BlankLine
    		Prio. queue of paths $Q \leftarrow{} \emptyset{}$\label{algo:mda*:initStart}\tcp*{Lex. non-decreasing order w.r.t. $\gamma$.}
    		Transition graph $\transitionGraph$ initialized only containing the transition node $\{s\}$\label{algo:igmda:implicitInit}\;
    		Trivial $\{s\}$-$\{s\}$-path $p_{\text{init}} \leftarrow{} ()$\;
    		$Q \leftarrow{} Q.\mathtt{insert}(p_{\text{init}})$\label{algo:mda*:insertInitialPath}\;
    		\BlankLine
    		\While{$Q \neq \emptyset$}{
    			$p \leftarrow{} Q.\mathtt{extractMin}()$ \label{algo:mda*:extraction}\;
    			$U \leftarrow{}$ last transition node of path $p$ \tcp*[r]{If $p = p_{\text{init}}$, $U \leftarrow \{s\}$.}
    			%$\bar{c}_{dr}(\P_{sv}) \leftarrow$ merge$\Big(\bar{c}_{dr}(\P_{sv}), \´bar{c}_{dr}(p)\Big)$\label{algo:mda*:merge}
    			\BlankLine
    			\texttt{success} $\leftarrow$ \texttt{False}\;
    			\lIf{$U \neq V$} {
    				%					$\mathcal{P}_{st} \leftarrow (\mathcal{P}_{st}, p)$\;
    				$(Q, \mathtt{success}) \leftarrow{}$ \refAlgo{propagate}$(p, Q, T^*, \NQP, \beta_t)$\label{algo:mda*:propagate}
    			}
    			\lIf{$U == V$ or $\mathtt{success} == \mathtt{True}$}{$\mcs{U}.\mathtt{append}(p)$\label{algo:mda*:linePushBack2}}
    			\BlankLine
    			New queue path $p'$ for $U$ $\leftarrow{}$ Solve \eqref{eq:nqp} according to $U$ and the existing $\NQP_a$ lists for $a \in \incoming{U}^*$ \label{algo:mda*:ncl}\;
    			\lIf{$p' \neq \Null$} 
    			{$Q.\mathtt{insert}(p')$\label{algo:mda*:insert}}
    		}
    		\Return $P_{V}$;
    		
    		\caption[caption]{Implicit Graph Multiobjective Dijkstra Algorithm (IG-MDA)}\label{algo:igmda}}
    \end{algorithm}
    
    Paths in the algorithm are partitioned into two disjoint classes.
    \begin{description}
    	\item[Permanent paths] For every node $U \in \transitionGraph$, the set $\mcs{U}$ stores efficient and non-equivalent $\{s\}$-$U$-paths that might be relevant for the reconstruction of efficient $\{s\}$-$V$-paths stored in the minimum complete set $\mcs{V}$ of efficient $\{s\}$-$V$-paths w.r.t. $\gamma$ at the end of the algorithm. Thus, a path might become permanent during the algorithm and once it is label as such, it is stored until the end.
    	\item[Explored paths] Consider a transition node $U \in \transitionNodes$ and an $\{s\}$-$U$-path $p$ in $\transitionGraph$ that the IG-MDA considers for the first time. $p$ is an explored path if it is neither dominated by nor equivalent to a permanent path in $\mcs{U}$. In later iterations, the expansions of $p$ along the outgoing arcs $(U,W) \in \outgoing{U}^*$ are considered. If at least one such expansion of $p$ turns out to be an explored $\{s\}$-$W$-path, $p$ is made permanent and stored in $\mcs{U}$. Otherwise, $p$ is discarded and never considered again. In any case, $p$ is no longer an explored path after its expansions are considered. The IG-MDA stores explored paths in different datastructures.
    	\begin{description}
    		\item[Priority Queue] A priority queue $Q$ contains at most one explored $\{s\}$-$U$-path for every $U \in \transitionNodes$ at any point in time. The $\{s\}$-$U$-path in $Q$, if there is any, is required to be a lex. smallest explored $\{s\}$-$U$-path.
    		
			\item[Next Queue Path ($\NQP$) Lists] Other explored $\{s\}$-$U$-paths, if there are any, are stored in so called $\NQP$ lists. There is such a list associated with every arc in $\transitionArcs$. If the last arc of an explored $\{s\}$-$U$-path $p$ that is not stored in $Q$ is the arc $a \in \transitionArcs$, $p$ is stored in the list $\NQP_a$. Every such list is also sorted in lex. non-decreasing order.
    	\end{description}
    \end{description}
    
    In every iteration of the IG-MDA, a lex. minimal explored path is extracted from the queue $Q$ (\Cref{algo:mda*:extraction}). Assume $p$ is an extracted $\{s\}$-$U$-path for some node $U \in \transitionNodes$. Since $p$ is an explored path, it is not dominated by or equivalent to any path in $\mcs{U}$ by definition. Since for any $U \in \transitionNodes$ the explored path in $Q$ is lex. minimal compared to other existing explored $\{s\}$-$U$-paths, $p$ is guaranteed to be an efficient $\{s\}$-$U$-path in $\transitionGraph$ w.r.t. $\gamma$. Thus, when extracted from $Q$, the IG-MDA builds the expansions of $p$ along the outgoing arcs in $\outgoing{U}^*$ to decide whether $p$ is made permanent or can be discarded.
    
    \begin{algorithm}
    	\small{
    		\SetKwInOut{Input}{Input}
    		\SetKwInOut{Output}{Output}
    		\Input{$\{s\}$-$U$-path $p$, priority queue $Q$, permanent paths $T^*$, lists of explored paths $\NQP$.}
    		\Output{Updated priority queue $Q$ and a boolean flag telling if the propagation of $p$ was successful along at least arc in $\outgoing{U}^*$.}
    		\BlankLine
    		Flag $\mathtt{success} \leftarrow $ False\;
    		\lIf{$\outgoing{U}^*$ not initialized}{build $\outgoing{U}^*$ as described in \Cref{sec:implicitNodes} \label{algo:propagate:newArcs}}
    		\For{$W \in \outgoing{U}^*$}{
    			$q \leftarrow p \circ (U,W)$\;
    			\lIf{$\gamma(\mcs{W}) \dom \gamma(q)$}{\textbf{continue}}
    			$\mathtt{success} \leftarrow $ True\;
    			\uIf{$Q$ does not contain an $\{s\}$-$W$-path}{
    				$Q.\mathtt{insert}\left( q \right)$ \label{algo:propagate:insert}\;
    			}
    			\uElse{
    				$q' \leftarrow \ \{s\}$-$W$-path in $Q$\;
    				\If{$\gamma(q) \lexsmaller \gamma(q')$ \label{algo:propagate:flexCheck}}{
    					$Q.\mathtt{decreaseKey}(W, \, q)$ \label{algo:propagate:decrease}\;
    					
    					\If{\textbf{not} $\gamma(q) \leq \gamma(q')$} {
    						$(U',W) \leftarrow$ last arc in path $q'$\;
    						Insert $q'$ at the beginning of $\NQP_{(U', W)}$\;\label{algo:propagate:pushFront}
    					}
    				}
    				\Else{
    					\lIf {\textbf{not} $\gamma(q') \leq \gamma(q)$}{
    						Insert $q$ at the end of $\NQP_{(U,W)}$\label{algo:propagate:pushBack}
    					}
    				}
    			}
    		}
    		\Return $(Q, \mathtt{success})$\;
    		\caption{propagate.}\label{algo:propagate}
    	}
    \end{algorithm}
    
    The decision is made in the subroutine \refAlgo{propagate} called in \Cref{algo:mda*:propagate} of the IG-MDA. Consider an arc $(U,W) \in \outgoing{U}^*$. If the new explored $\{s\}$-$W$-path $q := p \circ (U,W)$ is neither dominated by nor equivalent to a path in $\mcs{W}$, $q$ becomes an explored path and $p$ is stored as a permanent path in $\mcs{U}$. If no expansion of $p$ along an outgoing arc in $\outgoing{U}^*$ becomes a new explored path, $p$ cannot be expanded to become an efficient $\{s\}$-$V$-path in $\transitionGraph$ and is thus not needed and neglected. For every successful expansion of $p$ along an outgoing arc $a = (U, W) \in \outgoing{U}$, the new explored $\{s\}$-$W$-path $q$ is stored as a new explored  path. If there is no $\{s\}$-$W$-path in $Q$, $q$ is inserted into $Q$ (\Cref{algo:propagate:insert}). Otherwise, we assume $q'$ to be the $\{s\}$-$W$-path in $Q$ and distinguish two cases.
    If $\gamma(q') \lexsmaller \gamma(q)$ and $q$ is neither dominated nor cost equivalent to $q'$, $q$ is stored at the beginning of the list $\NQP_a$ of its last arc $a \in \transitionArcs$ \Cref{algo:propagate:pushBack}. If $\gamma(q) \lexsmaller \gamma(q')$, $q$ substitutes $q'$ in $Q$ and $Q$ is resorted. $q'$ might re-enter the priority queue later since it can still become a permanent $\{s\}$-$W$-path. Thus, if $q'$ is neither dominated by nor cost equivalent to $q$, it is stored in the list $\NQP_{a'}$ of its last arc $a' \in \transitionArcs$ \Cref{algo:propagate:pushFront}. Inserting at the front or at the back of the $\NQP$ lists depending on the case is explained in \citep[Lemma 4]{Casas22}. In both cases, the lists remain sorted in lex. non-decreasing order.
    
    Besides the propagation of $p$ as described in the last paragraph, every iteration of the IG-MDA needs to search for a new explored $\{s\}$-$U$-path to be stored in the priority queue $Q$. Recall that for every node $U$, there is at most one $\{s\}$-$U$-path in $Q$ at any point in time and that this path, if it exists, is lex. minimal among all explored $\{s\}$-$U$-paths. Thus, after the extraction of $p$ from $Q$, a new explored $\{s\}$-$U$-path $p'$ might be inserted into $Q$. The explored $\{s\}$-$U$-paths are stored in the $\NQP_a$ lists for $a \in \incoming{U}^*$ and thus, $p'$  solves the minimization problem
    \begin{equation}
    	\label{eq:nqp}
    	\arglexmin \Big\{ \gamma(p) \Big| p \in \NQP_a,\, a \in \incoming{U}^*, \text{ and not } \gamma(\mcs{U}) \dom \gamma(p) \Big\}.
    \end{equation}
    The minimization and the addition of $p'$ to $Q$ in case such a new explored path is found happen in \Cref{algo:mda*:ncl} and \Cref{algo:mda*:insert} of \Cref{algo:igmda}.
    
	Further details on how the original T-MDA proceeds, its correctness, and further speedup techniques in the implementation can be read in \citep{Casas22}. We now explain how $\transitionGraph$ is build and handled in the IG-MDA.
	
	\subsection{Implicit Handling of the Transition Graph}
    \label{sec:implicitNodes}

    In this section we exlpain how the transition graph $\transitionGraph$ is managed implicitely in the IG-MDA. 
    The addition of a node $U$ to the initially empty set $\transitionNodes$ of transition nodes always entails the initialization of the list $\mcs{U}$ since we assume that the algorithm will store some permanent $\{s\}$-$U$-paths. No arcs are added during this node initialization. At the beginning of the algorithm, the IG-MDA only adds the node $\{s\}$ to $\transitionGraph$ (\Cref{algo:igmda:implicitInit}).

    When an $\{s\}$-$U$-path $p$ is extracted from the priority queue, expansions of $p$ along outgoing arcs of $U$ are built in \refAlgo{propagate}. To this aim, the set $\outgoing{U}^*$ needs to be constructed if it does not yet exist (\Cref{algo:propagate:newArcs} of \refAlgo{propagate}). The construction happens using \eqref{eq:outgoingArcs} and the arc removal techniques discussed in \Cref{lem:efficientParallelArcs} and \Cref{lem:efficientCutExit}. The creation of an arc $(U,W)$ in this set requires
    \begin{itemize}
        \item the addition of $(U,W)$ to the set $\outgoing{U}^*$,
        \item the addition of $W$ to $\transitionNodes$ as explained in the last paragraph, if it does not yet exist, 
        \item the addition of $(U,W)$ to $\incoming{W}^*$,
        \item and the initialization of the list $\NQP_{(U,W)}$
    \end{itemize}
    In case the set $\outgoing{U}^*$ already exists when $p$ is extracted from the queue, we are sure that the lists $\NQP_{(U,W)}$ that are possibly needed in \refAlgo{propagate} are already initialized and that the new explored paths $q = p \circ (U,W)$ obtained from $p$'s expansions can be made permanent, i.e., added to $\mcs{W}$, if needed later.

    \subsection{Running Time}
    Using the running time bound for the T-MDA derived in \citep{Maristany21}, the number of nodes \eqref{eq:transitionNodesCount}, and the number of arcs \eqref{eq:transitionArcsCount} in $\transitionGraph$, we obtain the following result.
    
    \begin{theorem}
        Consider an  MO-MST instance $(G,c,s)$ and the corresponding One-to-One MOSP instance $\mathcal{I}_{\text{SP}} = (\transitionGraph^*, \gamma, \{s\}, V)$. Using the IG-MDA to solve $\mathcal{I}_{\text{SP}}$, setting $N_{\max} := \max_{U \in \transitionNodes} |\mcs{U}|$, and applying \Cref{thm:treePathEquiv}, $(G,c,s)$ can be solved in 
        \begin{equation*}
            \bigoh{d N_{\max} \Big(|\transitionNodes| \log{|\transitionNodes|} + |\transitionArcs|N_{\max} \Big) }.
        \end{equation*}
    \end{theorem}
    
    \section{Experiments}\label{sec:experiments}

    All codes, results, and evaluation scripts used to generate the contents in this section are publicly available in \citep{code}.

    \subsection{Implementation Details} 
    \label{sec:implementationDetails}
    As noted in the introduction, when choosing a benchmark algorithm for our experiments in \Cref{sec:experiments}, we benefit from the extensive survey on MO-MST algorithms by \citet{Fernandes2020}. For more than two objectives, our use case, the best results are obtained using the \emph{Built Network} (BN) algorithm, a dynamic programming algorithm originally published in \citet{Santos18}. We explain the details of our implementation of the BN algorithm and assess its efficiency in \ref{sec:bn}. 
    
    The deletion of red and blue edges from the original input graphs as explained in \Cref{sec:polyTimeRed} is done in a preprocessing stage and both algorithms work on the resulting graphs without red or blue edges. In \citep[Table 13]{Fernandes2020} the authors list the number of red and blue edges in many MO-MST instances that we use in our experiments. In our code, we do no further manipulation of the input data.
    
    A technique used by \citet{Pulido14} to speedup the $\dom$ dominance checks is called \emph{dimensionality reduction}. We already used it in \citep{Casas22} to design the T-MDA and obtained a considerable speedup. Dimensionality reduction is used in conjunction with lexicographic sorting of explored paths. This sorting implies $\gamma_1(p) \leq \gamma_1(p')$ if a path $p'$ is extracted from the queue after a path $p$. Thus, to determine whether $p'$ is dominated by $p$, it suffices to compare the remaining $(d-1)$ cost dimensions. Further implications of this technique are discussed in detail in \citep{Pulido14, Casas22}. Note that in \citep{Santos18} the authors achieved their fastest running times  using the \emph{sum of cost components} as the sorting criteria for explored paths. Also in our experiments this criteria yields better running times than lex. sorting. However, adding the dimensionality reduction technique to the BN algorithm when using lex. sorting, the overall running time is clearly better than using the sum of cost components as sorting criteria. Therefore, we report the results obtained from our BN implementation using lex. sorting and the dimensionality reduction technique.

    Further implementation details such as the representation of paths using labels and the usage of a memory pool to avoid memory fragmentation can be read in \citep{Casas22} or in our source code \citep{code} directly. The running time of our implementation of the BN algorithm is dominated by the computation of minimal paths (cf. \Cref{def:minimalPaths}). Thus, it is worth noting that for every explored path, we store a vector that indicates the order in which nodes of $G$ are added to the corresponding tree. We then use this vector in \Cref{algo:bn:minPaths} of \Cref{algo:bn} to be able to recognize relevant outgoing arcs faster in every iteration.

    Readers interested in further implementation details will notice that in \citep{Casas22} the original T-MDA was designed for One-to-One MOSP problems in which a lower bound, also called heuristic, of the costs from any (transition) node to the target (transition) node is computed before the algorithm starts. However, since the transition graphs used in this paper for the MOSP calculations are implicit, we cannot derive good lower bounds in advance.  Thus, while reading \citep{Casas22} always assume that our lower bound on the efficient paths' costs is the zero vector.

    \subsection{Instance Description}
    \label{sec:instances}

    \begin{description}
        \item[SPACYC based instances from \citep{Santos18}] The authors from \citep{Santos18} kindly gave us access to their instances from which we use the $3$ and $4$ MO-MST instances for our experiments. For each number of criteria, graphs with $5$ to $14$ nodes are generated. For each fixed number of $n$ nodes, graphs with $m = n i$ edges for $i \in \{5, 10, 15, 20\}$ are generated. For each arc, the costs are generated randomly using the \emph{SPACYC} generator from \citep{Knowles01}. Thereby, the cost criteria are not correlated and the costs are chosen from the interval $[0,100]$.
        \item[Instances from \citep{Fernandes2020}] We also got access to the instances used in \citet{Fernandes2020}. In this paper, we use their $3$ and $4$ dimensional MO-MST instances. For each dimension, we consider grid graphs and complete graphs with varying number of nodes (see \citep{Fernandes2020, code} for details). Additionally, instances were generated with anticorrelated and correlated edge costs. For a fixed problem dimension ($3$ or $4$ cost criteria), a fixed number of nodes, and an edge costs type there are $30$ different instances. Thus, for example, we consider $30$ complete graphs with $12$ nodes and anticorrelated edge costs functions with $3$ cost criteria.
    \end{description}

    \subsection{Computational Environment}
    All computations were run on a machine with an Intel Xeon Gold $6246$ @ $3.30$GHz processor. The source code, written in C++, was compiled using g++ v.$7.5.0$ using the compiler flag $-O3$. Both algorithms were granted $2$h of time and $30$GB of memory to solve each instance. 
    
    Indexing transition nodes depending on the node set from the original graph $G$ that they represent is important in both algorithms. To achieve this, we use the \texttt{dynamic bitset} class from the \texttt{boost} library \citep{BoostxLibrary} and for every bitset of length $n-1$ (since the node $s$ is fixed) we compute the corresponding decimal representation to obtain an index. By doing so, we restrict ourselves to graphs with at most $64$ nodes in today's $64$bit systems. However, as we will see in this section, the amount of efficient trees in every considered graph type requires more than the available $30$GB of memory before reaching our constraint on the number of nodes.

    \subsection{Results}
    In this section we finally report the results obtained from our comparison between the IG-MDA and the BN algorithm implemented in \citep{code}. Note for small input graphs, the obtained running times were below $0.01$s for both algorithms. In the tables in this section, we do not report results for any group of graphs for which both algorithms solved the instances in less than $0.01$s. However, in the \texttt{results} folder in \citep{code}, all results can be accessed and we also included the scripts used to generate the full tables and plots from this section. Also, for every table in the upcoming subsections, we pick a scatter plot corresponding to the running times of both algorithms that lead to one representative line of the table. The remaining scatter plots, one per line, are also stored in the \texttt{results} folder in \citep{code}.
    
    \subsubsection{Results from SPACYC based instances}

\begin{figure}
    		\begin{minipage}{.48\linewidth}
    			\captionsetup{type=figure}\includegraphics[width=\textwidth]{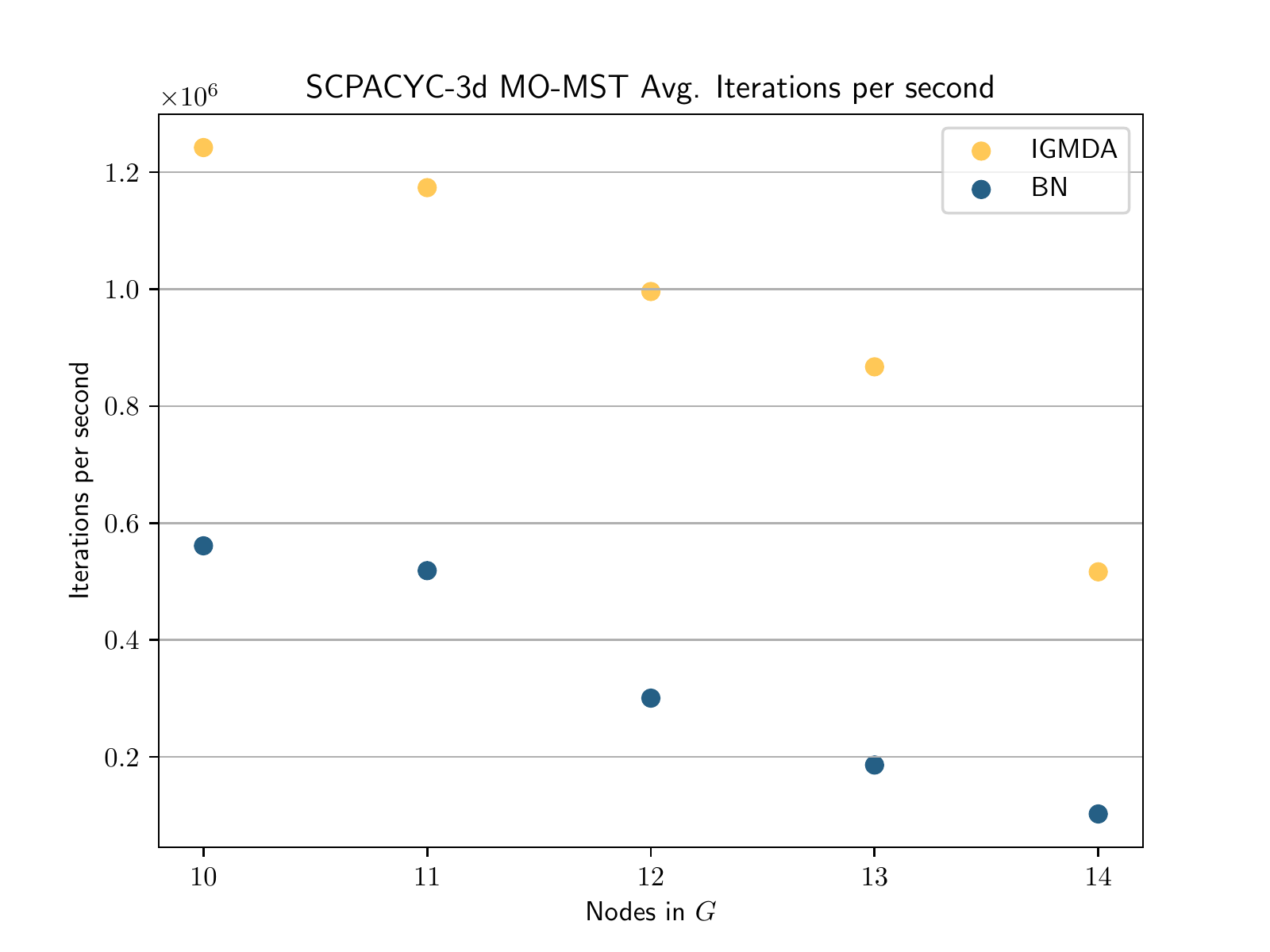}
    			\captionof{figure}{SPACYC $3$d, avg. iterations/second}\label{fig:SPACYC-3d-its/sec}
    		\end{minipage}
    		\begin{minipage}{.48\linewidth}
    			\captionsetup{type=figure}\includegraphics[width=\textwidth]{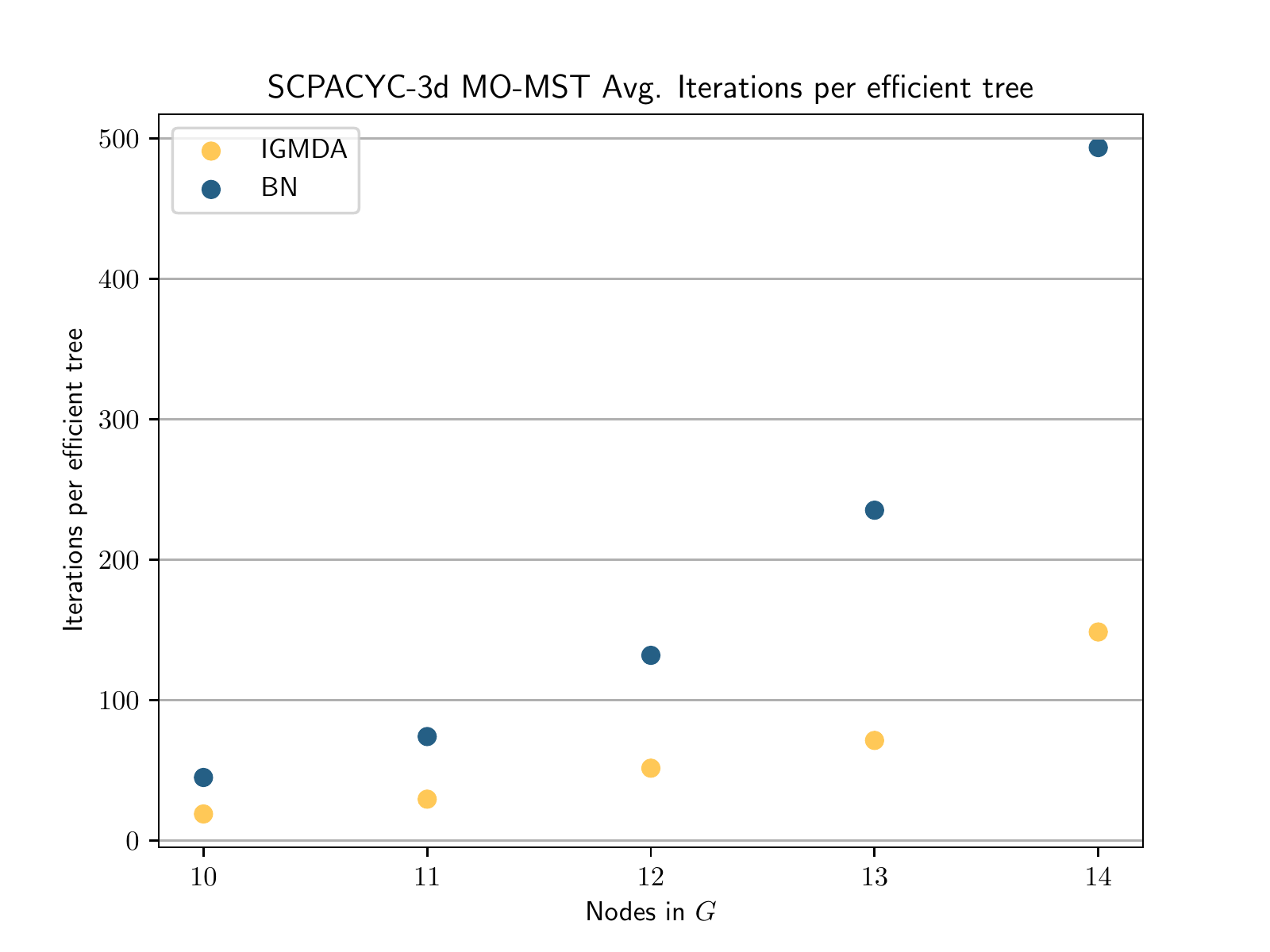}
    			\captionof{figure}{SPACYC $3$d, avg. iterations/solution}\label{fig:SPACYC-3d-its/sol}
    		\end{minipage}
    	\end{figure}

\begin{figure}
	    	\begin{minipage}{.48\linewidth}
	    		\captionsetup{type=figure}\includegraphics[width=\textwidth]{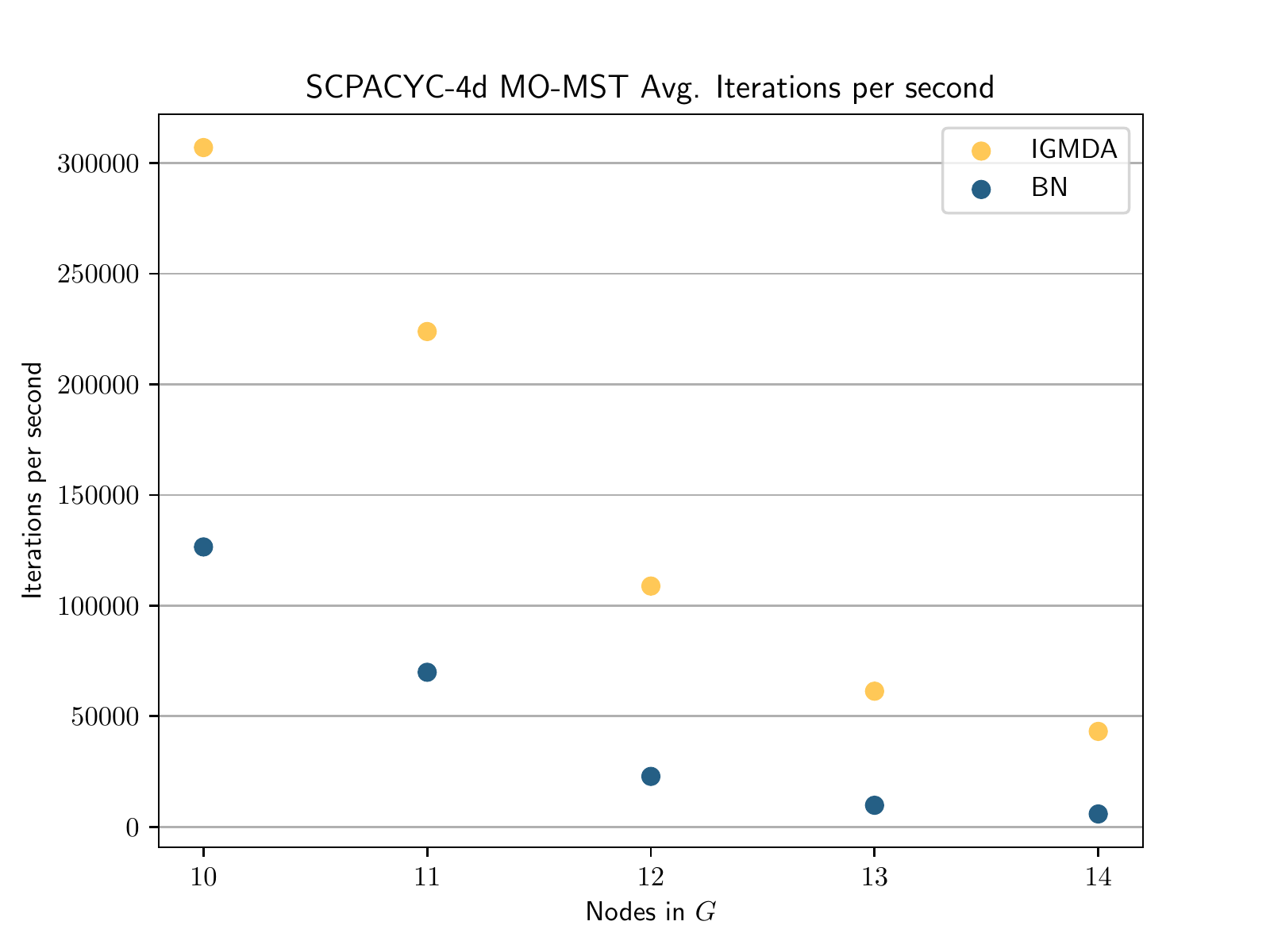}
	    		\captionof{figure}{SPACYC $4$d avg. iterations/second}\label{fig:SPACYC-4d-its/sec}
	    	\end{minipage}
	    	\begin{minipage}{.48\linewidth}
	    		\captionsetup{type=figure}\includegraphics[width=\textwidth]{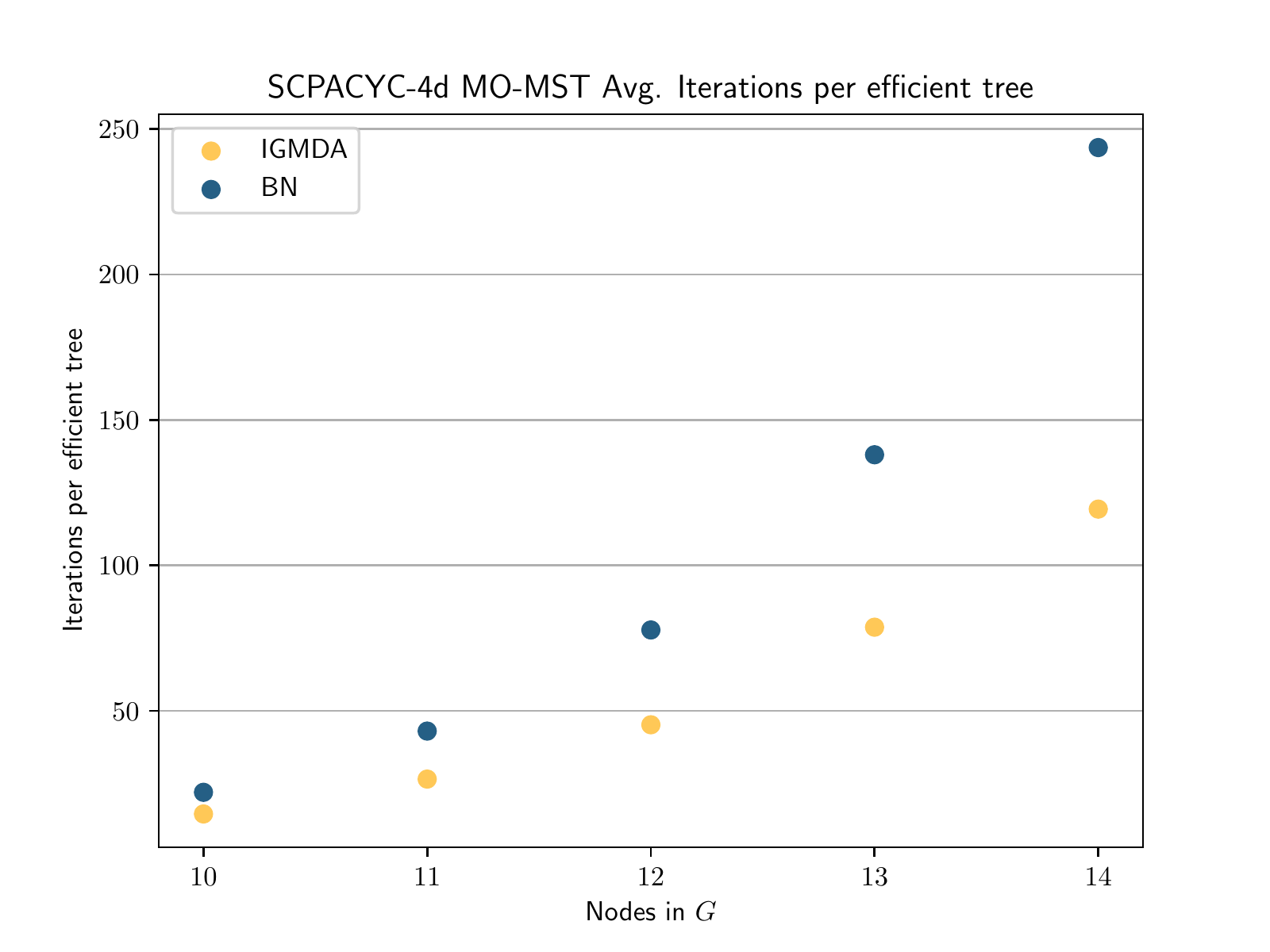}
	    		\captionof{figure}{SPACYC $4$d, avg. iterations/solution}\label{fig:SPACYC-4d-its/sol}
	    	\end{minipage}
    	\end{figure}

    In \Cref{tab:SPACYC} we summarize the results obtained from the SPACYC based instances. Both algorithms solve every instance. While in $3$d the BN algorithm is faster than the IG-MDA on instances with less than $8$ nodes, this effect disappears in $4$d, where the IG-MDA outperforms the BN algorithm consistently for all graph sizes. In both dimensions, the speedup grows with the input graph size and reaches $\times 16.76$ on the $3$d instances and $\times 15.04$ on the $4$d instances. Thereby, the IG-MDA performs better regarding the metrics \emph{iterations per second} and \emph{iterations per efficient spanning tree}. The second metric also implies that the IG-MDA solves the instances in less iterations since the number of efficient spanning trees depends on the instances and not on the algorithm. \Cref{fig:SPACYC-3d-its/sec} and \Cref{fig:SPACYC-4d-its/sec} show, for graphs with $10$ to $14$ nodes, the number of iterations that the IG-MDA and the BN algorithm performed per second. The IG-MDA outperforms the BN algorithm regarding this metric consistently. Particularly in \Cref{fig:SPACYC-4d-its/sec} we observe that both algorithms seem to converge. This is because the dominance checks ($\dom$) become more complex as the number of efficient solutions in the sets $\mcs{U}$ for transition nodes $U$ increase. Since these sets are equal for both algorithms, both algorithms make the same effort to decide whether explored trees are dominated. If the iterations per second of both algorithms converge but the speedup in favor of the IG-MDA increases with the graph size, the IG-MDA must do less iterations than the BN algorithm. Indeed, this can be already observed in \Cref{tab:SPACYC}. In \Cref{fig:SPACYC-3d-its/sol} and \Cref{fig:SPACYC-4d-its/sol} we plot the iterations per efficient spanning tree needed by both algorithms. It becomes apparent regarding this metric, the effort made by the IG-MDA increases much slower as the graph size increases. Using \emph{minimal paths} (cf. \Cref{def:minimalPaths}), the BN algorithm needs to decide upon the relevance of new explored paths using dominance checks only. Using cost-dependent arc pruning techniques as described in \Cref{sec:domParallelArcs} and \Cref{sec:chen}, the IG-MDA avoids the expansion of every efficient $\{s\}$-$U$-path in $\mcs{U}$ along a pruned arc $(U,W)$ without doing dominance checks for every such path. In particular, being able to prune all incoming arcs of a transition node $W$ leads to less transition nodes being considered in the implicit graphs of the IG-MDA algorithm (cf. $|\transitionNodes|$ columns in \Cref{tab:SPACYC}).

        \begin{table}
            \scriptsize{
            \centering
            \caption{SPACYC based $3$d and $4$d instances. For every considered node cardinality $|V|$, $20$ instances were considered. Both algorithms solved every instance. Numbers are geometric means.}
            \label{tab:SPACYC}
            \begin{tabular}{R R R R R R R R R}
            	\toprule
                 \multirow{2}{*}{$|V|$} & \multirow{2}{*}{$|T_{G}^*|$} & \multicolumn{3}{c}{BN} & \multicolumn{3}{c}{IG MDA} & \multirow{2}{*}{Speedup} \\
                 \cmidrule(lr){3-5} \cmidrule(lr){6-8}
                & & |\transitionNodes| & \text{Iterations} & \text{Time} & |\transitionNodes| & \text{Iterations} & \text{Time} & \\
    			\midrule
				\multicolumn{9}{c}{$3$d edge costs} \\
				\midrule
% 5 & 14.95 & 12.49 & 29.94 & 0.0000 & 11.03 & 38.28 & 0.0000 & 0.69 \\
% 6 & 38.92 & 24.17 & 127.56 & 0.0001 & 22.03 & 137.10 & 0.0001 & 0.80 \\
% 7 & 75.35 & 51.35 & 447.73 & 0.0004 & 43.00 & 365.11 & 0.0004 & 0.98 \\
% 8 & 118.55 & 90.62 & 1196.81 & 0.0012 & 68.82 & 811.24 & 0.0006 & 2.11 \\
% 9 & 242.26 & 212.56 & 5015.96 & 0.0059 & 151.26 & 2755.71 & 0.0023 & 2.53 \\
10 & 398.14 & 477.71 & 17846.60 & 0.0318 & 312.88 & 7496.75 & 0.0060 & 5.27 \\
11 & 487.21 & 774.66 & 36049.40 & 0.0695 & 526.56 & 14310.60 & 0.0122 & 5.70 \\
12 & 815.86 & 1609.97 & 107572.10 & 0.3579 & 1179.25 & 41966.32 & 0.0421 & 8.50 \\
13 & 1073.61 & 3111.80 & 252588.01 & 1.3572 & 1789.97 & 76472.73 & 0.0882 & 15.39 \\
14 & 1747.29 & 6835.16 & 862549.99 & 8.4189 & 4288.41 & 259474.39 & 0.5023 & 16.76 \\
                    %\bottomrule
                 \midrule
                \multicolumn{9}{c}{$4$d edge costs} \\
                \midrule
                % 5 & 26.59 & 13.75 & 40.72 & 0.0 & 12.72 & 61.0 & 0.0 & 1.15 \\
                % 6 & 93.75 & 31.23 & 255.79 & 0.0 & 28.19 & 278.7 & 0.0 & 1.44 \\
                % 7 & 238.55 & 58.38 & 1045.39 & 0.0 & 53.69 & 1019.22 & 0.0 & 1.54 \\
9 & 1146.20 & 241.41 & 15941.44 & 0.0461 & 197.35 & 10460.27 & 0.0163 & 2.83 \\
10 & 3014.38 & 471.18 & 66369.32 & 0.5246 & 413.88 & 43938.54 & 0.1431 & 3.66 \\
11 & 4532.14 & 996.44 & 195159.07 & 2.7925 & 859.96 & 120389.30 & 0.5378 & 5.19 \\
12 & 9060.48 & 2002.31 & 705053.83 & 30.8805 & 1705.53 & 409772.59 & 3.7650 & 8.20 \\
13 & 13471.15 & 3734.22 & 1859534.46 & 190.0873 & 3196.62 & 1061076.64 & 17.2987 & 10.99 \\
14 & 20735.06 & 7470.75 & 5051489.41 & 862.0660 & 6237.86 & 2474854.74 & 57.3346 & 15.04 \\
                \bottomrule
            \end{tabular}
            }
        \end{table}
    
         \begin{figure}
            \begin{minipage}{.48\linewidth}
                \captionsetup{type=figure}\includegraphics[width=\textwidth]{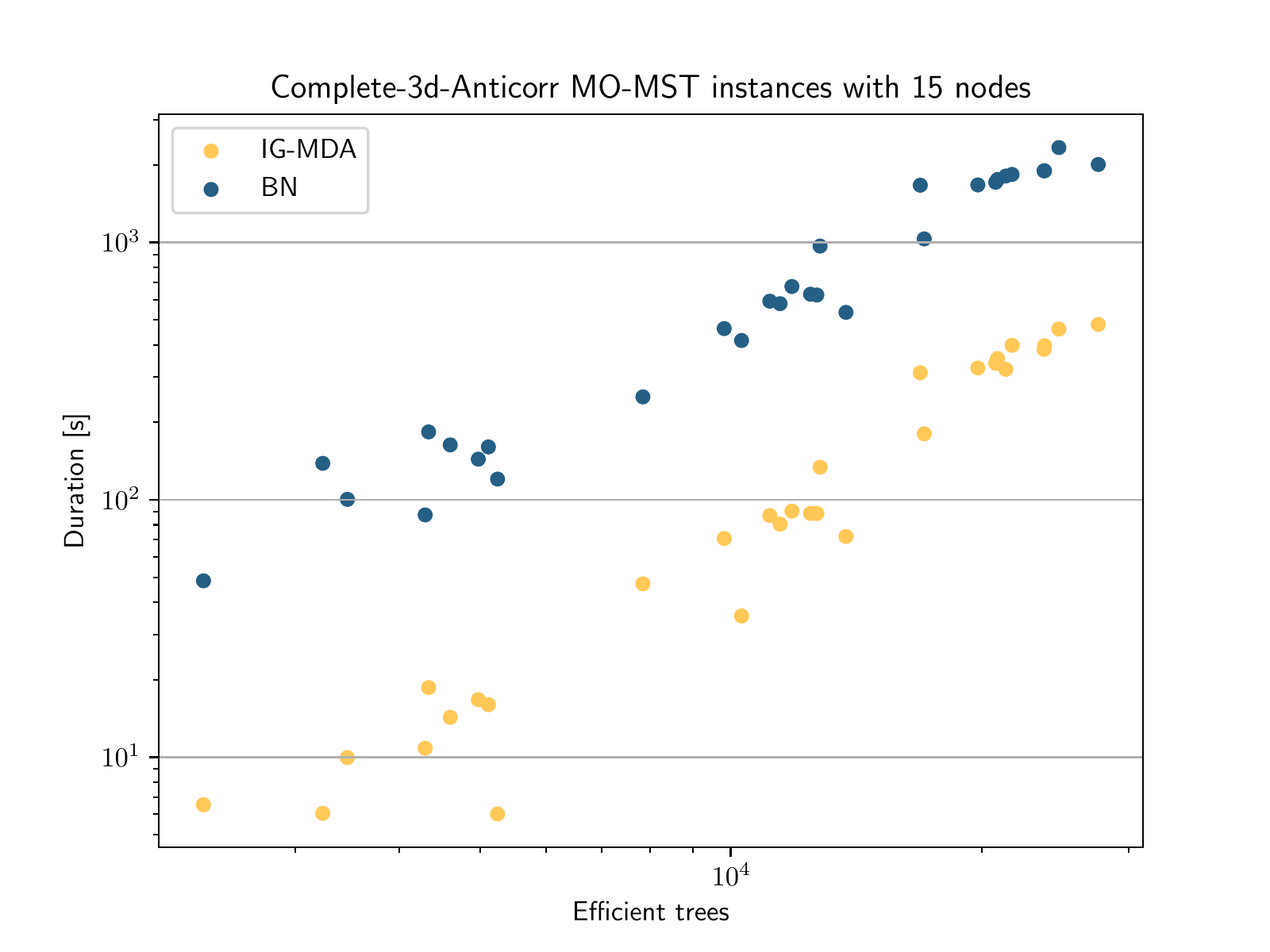}
                \captionof{figure}{Complete-$3$d-Anticorr, $15$ nodes}\label{fig:complete-3d-anticorr-15}
            \end{minipage}
            \begin{minipage}{.48\linewidth}
                \captionsetup{type=figure}\includegraphics[width=\textwidth]{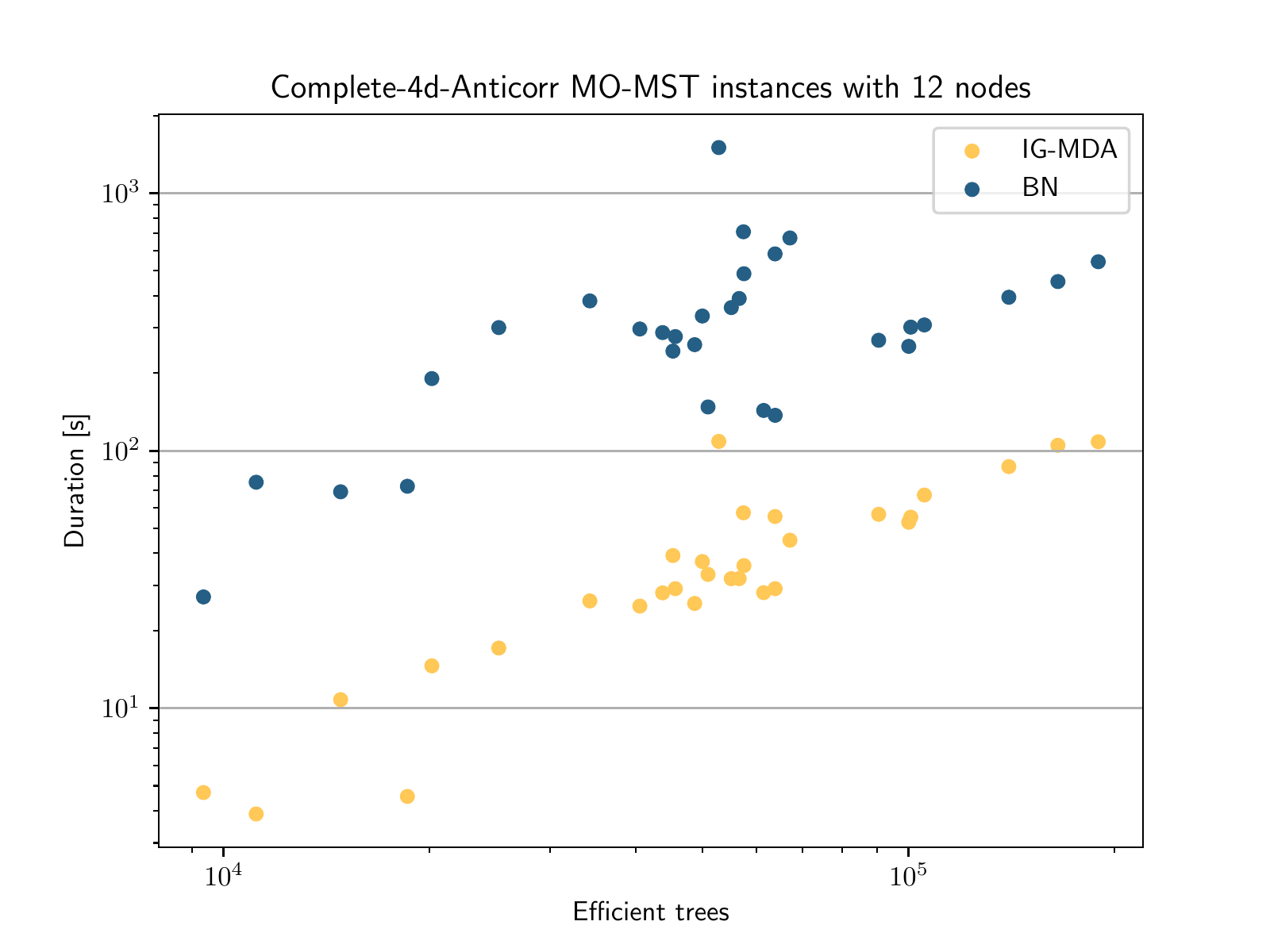}
                \captionof{figure}{Complete-$4$d-Anticorr, $12$ nodes}\label{fig:complete-4d-anticorr-12}
            \end{minipage}
        \end{figure}
    
        \subsubsection{Results from Fernandes et al. instances}
        \Cref{tab:anticorr} and \Cref{tab:corr} summarize the results obtained from the instances from \citep{Fernandes2020}. 
        
        \paragraph{Complete Graphs with Anticorrelated Costs} We report our results in \Cref{tab:anticorr}. This instances are the most difficult ones. Both algorithms solve all instances with up to $12$ nodes regardless of the edge cost dimension. Thereby, the IG-MDA is $\times 7.22$ faster on $3$d instances and $\times8.50$ faster on $4$d instances. Additionally, the IG-MDA solves all $3$d anticorrelated instances with $15$ nodes and all but one with $17$ nodes. Regarding the $4$d instances, it solves $22$ instances with $15$ nodes. This is an improvement w.r.t. to the biggest instances of this type solved in \citep{Fernandes2020} ($12$ nodes in $3$d and $10$ nodes in $4$d). Note that the time limit is not the bottleneck. Instead, computations are aborted because of the memory limit of $30$GB. \Cref{fig:complete-3d-anticorr-15} and \Cref{fig:complete-4d-anticorr-12} show the running times of both algorithms for some relevant graph sizes. The same plots for every other considered graph size are in the \texttt{result} folder from \citep{code}.

\begin{figure}
            \begin{minipage}{.48\linewidth}
                \captionsetup{type=figure}\includegraphics[width=\textwidth]{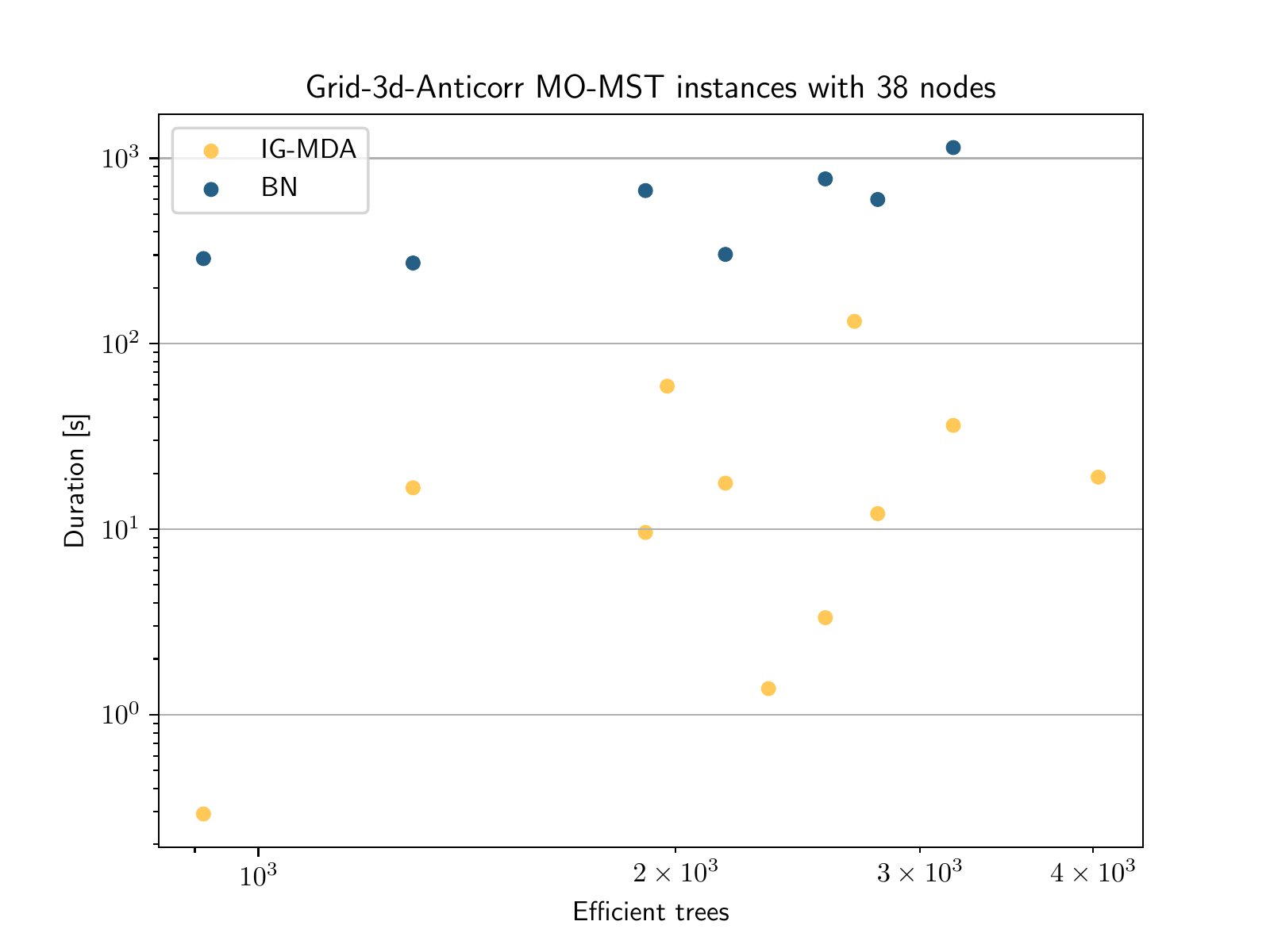}
                \captionof{figure}{Grid-$3$d-Anticorr, $38$ nodes}\label{fig:grid-3d-anticorr-38}
            \end{minipage}
            \begin{minipage}{.48\linewidth}
                \captionsetup{type=figure}\includegraphics[width=\textwidth]{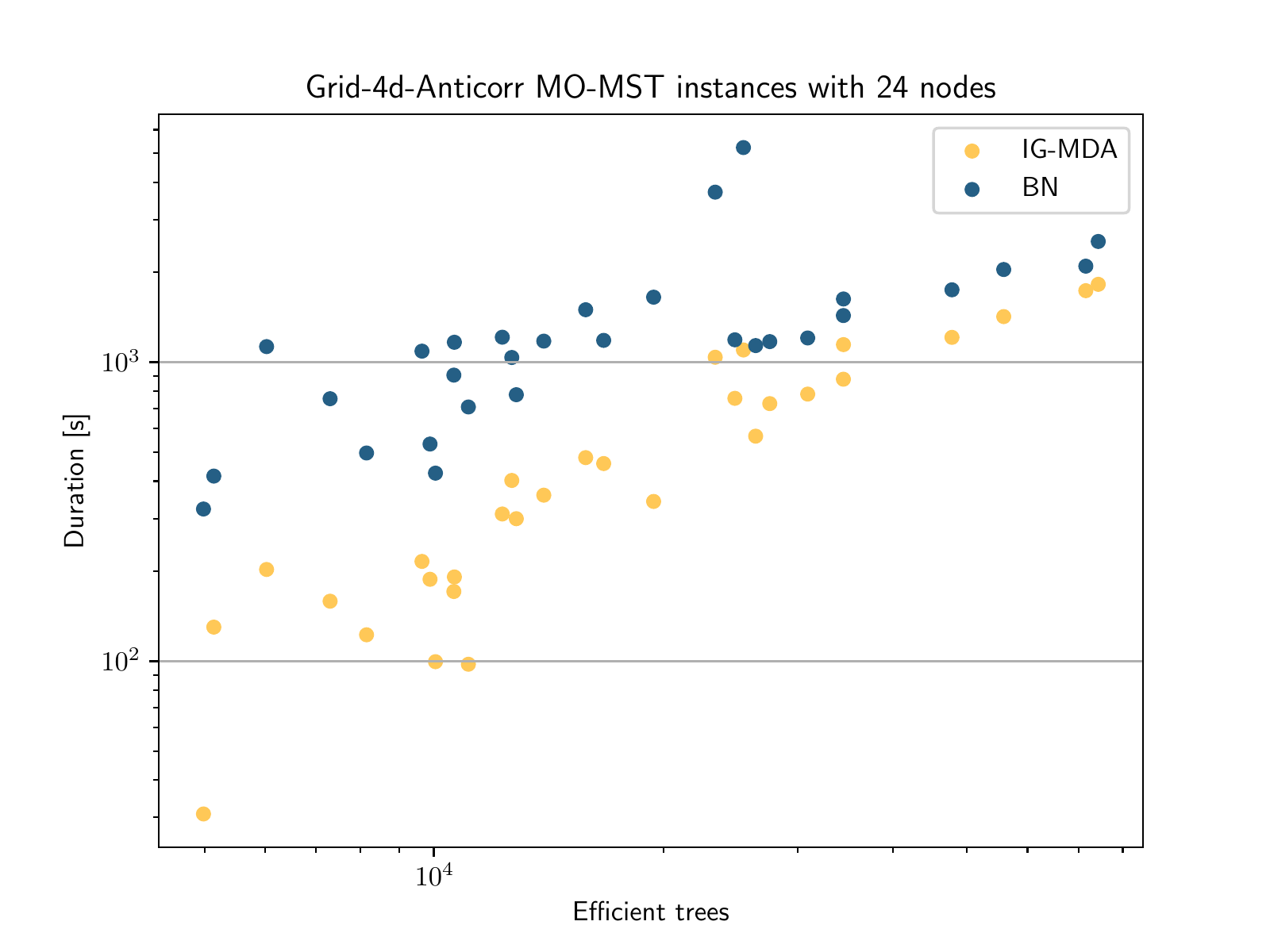}
                \captionof{figure}{Grid-$4$d-Anticorr, $24$ nodes}\label{fig:grid-4d-anticorr-24}
            \end{minipage}
        \end{figure}

        \paragraph{Grid Graphs with Anticorrelated Costs} The results are summarized in \Cref{tab:anticorr}. We observe the same trend as before regarding the speedups in favor of the IG-MDA. 
        All instances with graphs with up to $18$ nodes are solved in less than $0.01$s by both algorithms. On $3$d instances, both algorithms manage to solve all instances with up to $24$ nodes. On bigger graphs, the BN algorithm solves a subset of the solved by the IG-MDA. The IG-MDA starts failing to solve instances on input graphs with $33$ nodes. There are seven instances with $38$ nodes that are solved by both algorithms (the IG-MDA solves $11/30$ such instances) and on these instances the IG-MDA is $\times 68.16$ faster (cf. \Cref{fig:grid-3d-anticorr-38}). Note that the biggest instances from this group solved in \citep{Fernandes2020} were grids with $20$ nodes.
        The reason why the BN algorithm manages to solve some instances with $38$ nodes even though it could not solve any instance with $36$ nodes and only two with $33$ nodes is that the seven solved instances with $38$ nodes contain multiple blue and red edges. Thus, the actual input graphs for both algorithms are smaller (for details, see \citep{code}).
        Regarding instances with $4$d anticorrelated edge cost functions on grid graphs, both algorithms solve all instances with up to $24$ nodes. 
        The IG-MDA still solves all $30$ instances with $27$ nodes.
        From that size on the number of solved instances decreases because of the memory limit.
        The IG-MDA speedup on the instances with $24$ nodes is $\times 2.99$ (see \Cref{fig:grid-4d-anticorr-24}).
        The relatively small number compared to other instance sets is because grids induce smaller transition graphs and, particularly on $4$d instances, the running time is mostly determined by dominance checks ($\dom$) that are equally time consuming for both algorithms.
        %Since both algorithms compute the same sets $\mcs{U}$, their running times converge as the $\dom$ checks become more relevant.

        \begin{figure}
        \begin{minipage}{.48\linewidth}
            \captionsetup{type=figure}\includegraphics[width=\textwidth]{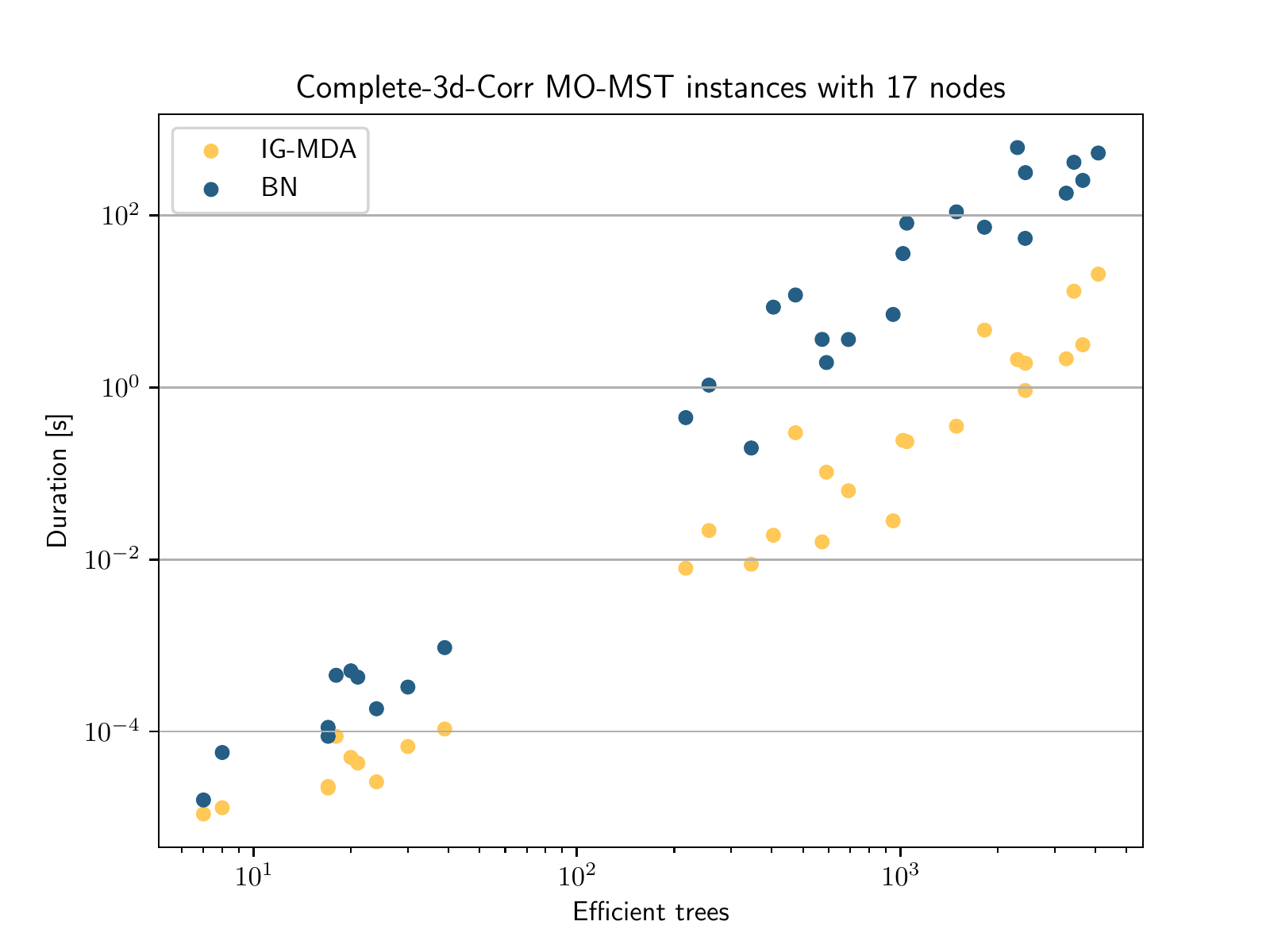}
            \captionof{figure}{Complete-$3$d-Corr, $17$ nodes}\label{fig:complete-3d-corr-17}
        \end{minipage}
        \begin{minipage}{.48\linewidth}
            \captionsetup{type=figure}\includegraphics[width=\textwidth]{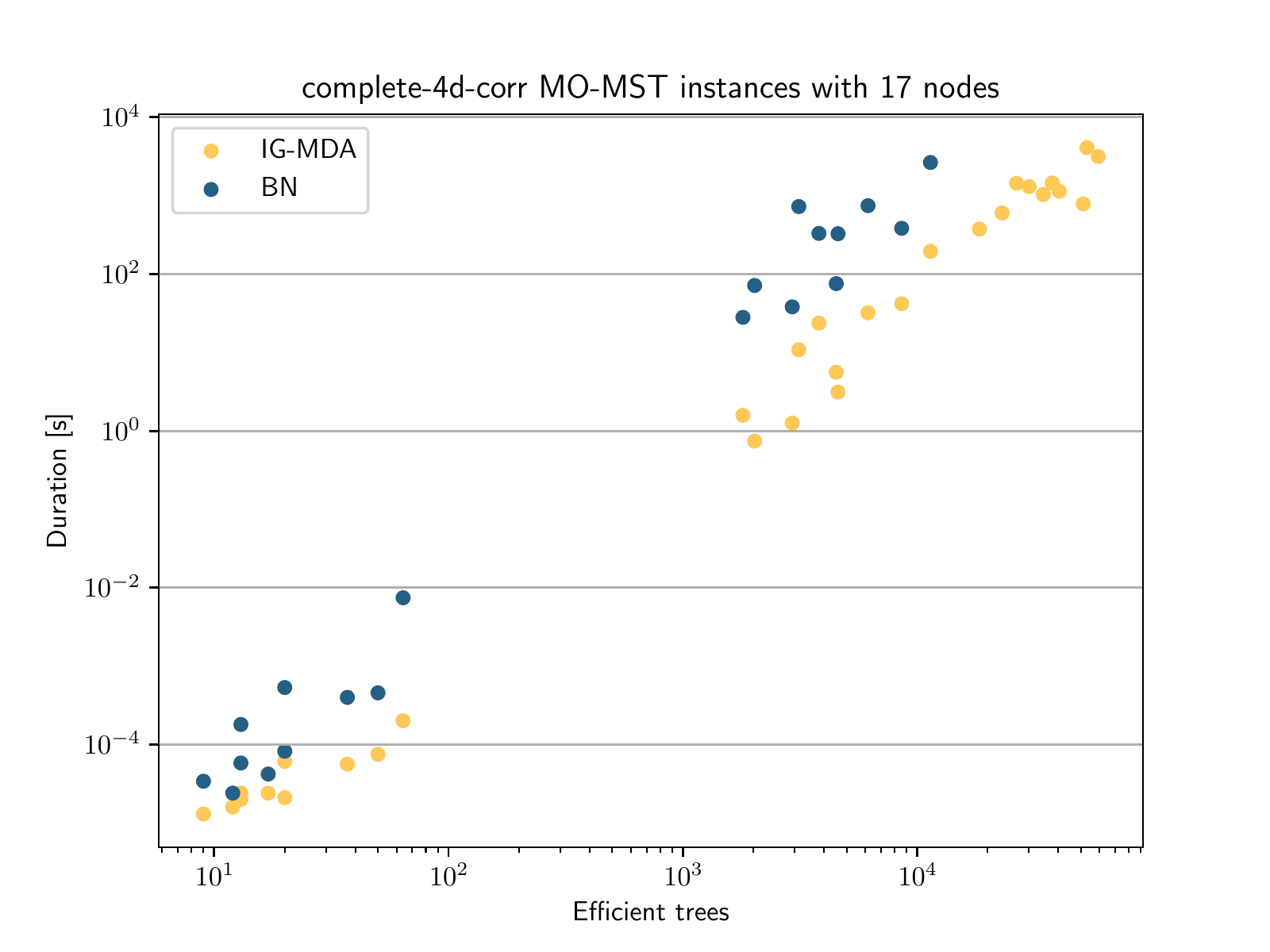}
            \captionof{figure}{Complete-$4$d-Corr, $17$ nodes}\label{fig:complete-4d-corr-17}
        \end{minipage}
    \end{figure}

    \begin{figure}
        \begin{minipage}{.48\linewidth}
            \captionsetup{type=figure}\includegraphics[width=\textwidth]{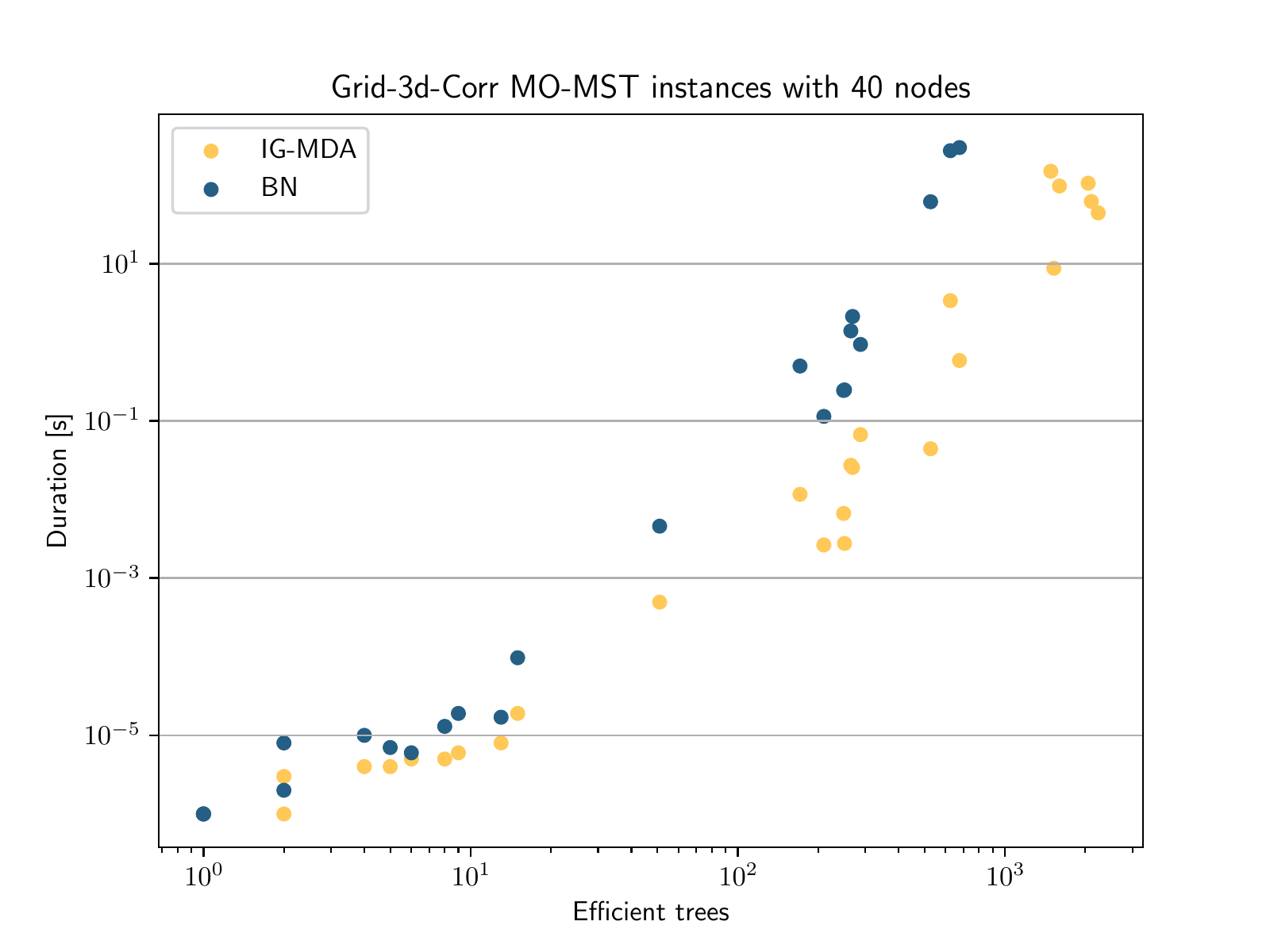}
            \captionof{figure}{Grid-$3$d-Corr, $40$ nodes}\label{fig:grid-3d-corr-40}
        \end{minipage}
        \begin{minipage}{.48\linewidth}
            \captionsetup{type=figure}\includegraphics[width=\textwidth]{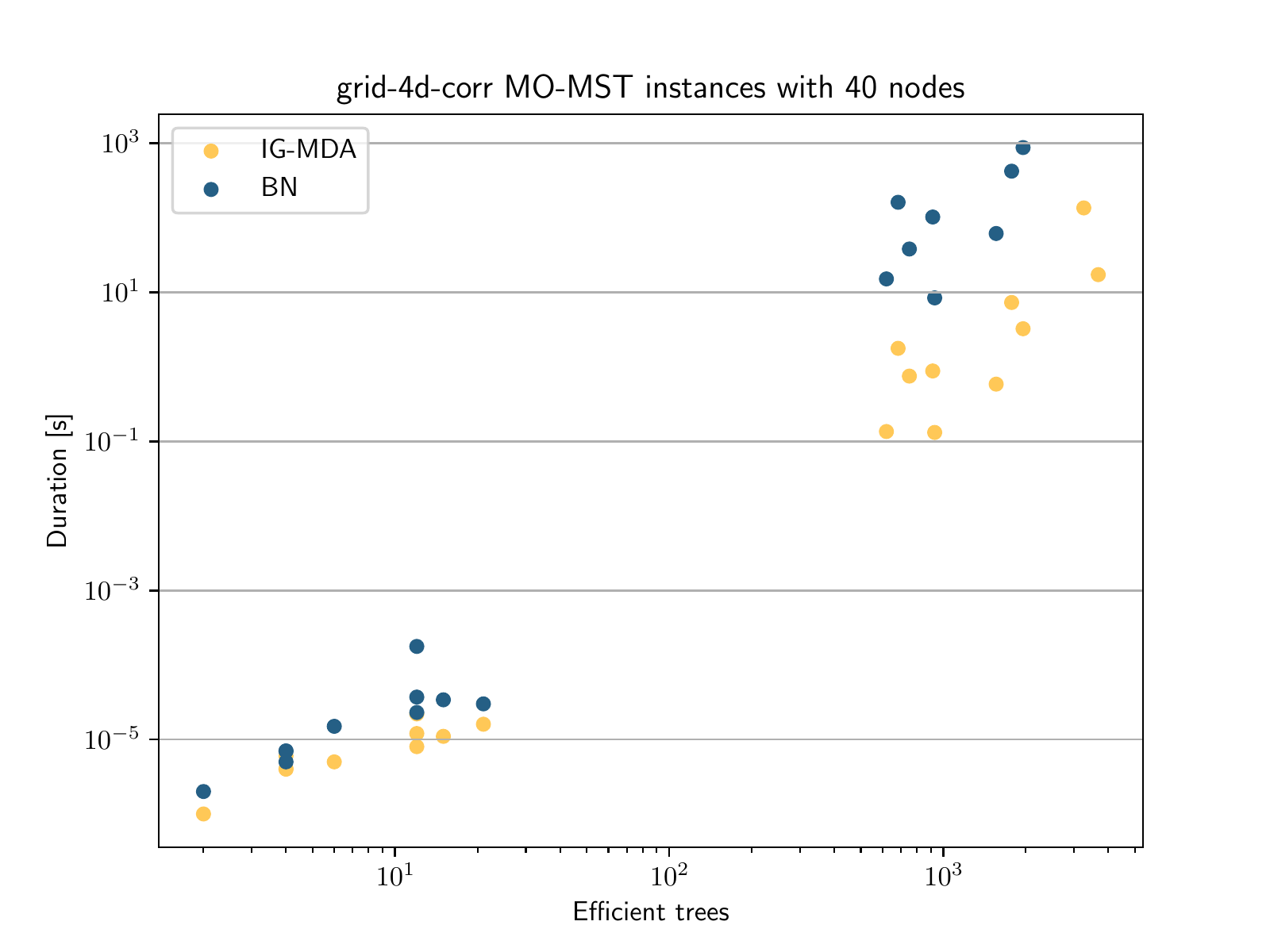}
            \captionof{figure}{Grid-$4$d-Corr, $40$ nodes}\label{fig:grid-4d-corr-40}
        \end{minipage}
    \end{figure}
        
        \begin{landscape}
        \begin{table}
            \scriptsize{
            \centering
            \caption{Instances from \citep{Fernandes2020} with anticorrelated edge costs. Cardinality $|T_G|$ of solution sets, $|\transitionNodes|$ of transition nodes, number of iterations, and time report geometric means built among solved instances only. The speedups consider only instances solved by both algorithms.}
            \label{tab:anticorr}
            \begin{tabular}{R R R R R R R R R R R R R}
                 \multirow{2}{*}{$|V|$} & \multirow{2}{*}{Insts} & \multicolumn{5}{c}{BN} & \multicolumn{5}{c}{IG MDA} & \multirow{2}{*}{Speedup} \\
                 \cmidrule(lr){3-7} \cmidrule(lr){8-12}
                & & Sol & |T_{G}^*| & |\transitionNodes| & \text{Iterations} & \text{Time} & Sol. & |T_{G}| & |\transitionNodes| & \text{Iterations} & \text{Time} & \\
                \midrule
               	\multicolumn{13}{c}{COMPLETE $3$d anticorr. edge costs}\\
                \midrule
                    %5 & 30 & 30 & 38.75 & 15.9 & 62.26 & 0.0 & 30 & 38.75 & 15.37 & 93.41 & 0.0 & 1.38 \\
                    %7 & 30 & 30 & 266.95 & 62.04 & 1484.21 & 0.0 & 30 & 266.95 & 59.61 & 1443.66 & 0.0 & 2.35 \\
                    10 & 30 & 30 & 1934.17 & 509.62 & 79059.14 & 0.4 & 30 & 1934.17 & 477.88 & 58042.67 & 0.08 & 4.82 \\
                    12 & 30 & 30 & 4042.89 & 2039.89 & 668348.06 & 7.68 & 30 & 4042.89 & 1868.86 & 434829.76 & 1.27 & 6.03 \\
                    15 & 30 & 30 & 10507.52 & 16344.5 & 13781596.11 & 530.79 & 30 & 10507.6 & 15166.2 & 8203636.87 & 73.52 & 7.22 \\
                    17 & 30 & 10 & 8995.82 & 65151.83 & 39926366.12 & 4545.6 & 29 & 16773.63 & 60128.77 & 50138867.03 & 775.91 & 10.22 \\

	    			\midrule
	    			\multicolumn{13}{c}{GRID $3$d anticorr. edge costs}\\
	    			\midrule
%6 & 30 & 30 & 9.87 & 13.37 & 19.13 & 0.00 & 30 & 9.87 & 13.01 & 28.43 & 0.00 & 1.10 \\
%12 & 30 & 30 & 84.68 & 163.27 & 923.40 & 0.00 & 30 & 84.68 & 126.34 & 786.87 & 0.00 & 1.81 \\
18 & 30 & 30 & 327.46 & 1970.48 & 27559.31 & 0.0444 & 30 & 327.46 & 1059.14 & 15957.42 & 0.0105 & 4.24 \\
20 & 30 & 30 & 507.80 & 4321.96 & 80646.78 & 0.1788 & 30 & 507.80 & 2081.46 & 41301.82 & 0.0354 & 5.05 \\
24 & 30 & 30 & 1905.73 & 289945.96 & 12561838.13 & 119.8732 & 30 & 1905.73 & 108148.81 & 4556890.58 & 19.8963 & 6.02 \\
27 & 30 & 25 & 2055.18 & 658592.00 & 32710242.90 & 331.8889 & 30 & 2432.92 & 243115.52 & 13189507.15 & 77.9690 & 7.27 \\
30 & 30 & 8 & 1400.97 & 798812.92 & 32607409.69 & 336.6812 & 20 & 2119.93 & 238092.28 & 13277318.30 & 75.8248 & 31.61 \\
33 & 30 & 2 & 2091.47 & 1353679.98 & 68100519.74 & 728.4824 & 16 & 3285.21 & 637610.27 & 44246349.41 & 341.4173 & 26.16 \\
36 & 30 & 0 & - & - & - & - & 10 & 2945.52 & 497406.55 & 34694216.43 & 263.5790 & - \\
38 & 30 & 7 & 1959.45 & 938641.43 & 51235269.92 & 504.1519 & 11 & 2189.33 & 35140.31 & 2977740.76 & 10.8611 & 68.16 \\
40 & 30 & 0 & - & - & - & - & 2 & 3440.77 & 1310244.04 & 98125888.60 & 874.0739 & - \\

                    \midrule
	    			\multicolumn{13}{c}{COMPLETE $4$d anticorr. edge costs}\\
					\midrule                    
                    %5 & 30 & 30 & 62.32 & 15.97 & 73.77 & 0.00 & 30 & 62.32 & 15.64 & 128.43 & 0.00 & 1.50 \\
                    %7 & 30 & 30 & 784.11 & 63.93 & 2798.93 & 0.00 & 30 & 784.11 & 63.43 & 3272.49 & 0.00 & 1.83 \\
                    10 & 30 & 30 & 14261.89 & 511.50 & 275211.78 & 3.61 & 30 & 14261.89 & 499.37 & 242326.48 & 0.85 & 4.27 \\
                    12 & 30 & 30 & 49937.31 & 2046.63 & 3635468.74 & 266.18 & 30 & 49937.32 & 2006.20 & 2951674.07 & 31.32 & 8.50 \\
                    15 & 30 & 0 & - & - & - & - & 22 & 200339.04 & 16224.95 & 77909983.48 & 4468.7791 & - \\
                    %17 & 30 & 0 & - & - & - & 7200.00 & 0 & - & - & - & 7200.00 & - \\
    	    			\midrule
    	    			\multicolumn{13}{c}{GRID $4$d anticorr. edge costs}\\
    	    			\midrule
						18 & 30 & 30 & 1605.54 & 3338.71 & 83063.06 & 0.1761 & 30 & 1605.54 & 2548.04 & 65344.60 & 0.0742 & 2.37 \\
						20 & 30 & 30 & 2022.36 & 6941.00 & 211043.31 & 0.5899 & 30 & 2022.36 & 4681.20 & 144836.59 & 0.2144 & 2.75 \\
						24 & 30 & 30 & 17036.10 & 487336.31 & 56591949.94 & 1145.0088 & 30 & 17036.10 & 308131.18 & 35528444.71 & 382.9531 & 2.99 \\
						27 & 30 & 12 & 10091.80 & 830056.49 & 70224110.93 & 1539.2769 & 30 & 21918.58 & 606263.90 & 83839892.33 & 1092.2398 & 5.93 \\
						30 & 30 & 0 & - & - & - & - & 8 & 21205.22 & 955448.68 & 133902354.68 & 2850.7280 & - \\
						33 & 30 & 0 & - & - & - & - & 2 & 13126.77 & 1776005.39 & 207538764.46 & 3671.7716 & - \\
     			\end{tabular}
     	}
    	\end{table}
        \end{landscape}

    The results from the instances with correlated edge cost functions are reported in \Cref{tab:corr}. As expected, the size of the solved instances is bigger than in the anticorrelated counterparts. However, this effect can be misleading. As we can observe in the scatter plots in \Cref{fig:complete-3d-corr-17} to \Cref{fig:grid-4d-corr-40} the running times of both algorithms are separated into two clusters. The reason is that with correlated costs, the preprocessing phase sometimes recognizes many blue and red arcs (cf. \Cref{sec:polyTimeRed}) s.t. the input graph for the MO-MST algorithms is actually very small. In fact, there are even $3$d grid instances with $36$ nodes that contain $35$ blue arcs s.t. the remaining graph consists only of one node. The $35$ blue arcs constitute the only efficient spanning tree for this instance. On grid graphs with a fixed number of nodes, the running times of both algorithm can differ by more than eight order of magnitudes (cf. \Cref{fig:grid-4d-corr-40}). Therefore, the geometric means in \Cref{tab:corr} are almost always clearly below one second but they need to be put into the context described in this paragraph. Only then, given the low average running times, we can understand that not all instances are solved.
    
    \paragraph{Complete Graphs with Correlated Costs}
    On $3$d instances on complete graphs (\Cref{tab:corr}), both algorithms solve all instances with up to $17$ nodes where the IG-MDA is $\times 29.09$ faster (\Cref{fig:complete-3d-corr-17}). The IG-MDA also solves all instances with $22$ nodes in $1.06$s on average. In this group of instances, the BN algorithm solves $19/30$ instances.
    Even though the edge costs are correlated, the instances on complete graphs with $4$d edge costs are difficult and the BN algorithm cannot solve all instances with $15$ nodes. 
    On these graphs, the speedup in favor of the IG-MDA is $\times 15$; higher than on the same graphs with $3$d edge costs.

    \paragraph{Grid Graphs with Correlated Costs} The lower left clusters in \Cref{fig:grid-3d-corr-40} and \Cref{fig:grid-4d-corr-40} show that indeed a considerable amount of instances in this group are almost solved during preprocessing. After deleting red edges and contracting blue ones, the remaining instances are so small that both MO-MST algorithms solve them in less than a millisecond. Overall, the IG-MDA remains consistently more than an order of magnitude faster than the BN algorithm on this type of instances (cf. \Cref{tab:corr}). After deleting red edges and contracting blue edges, we are left with $12$ instances with $4$d edge costs and $30$-$31$ nodes. To the best of our knowledge these instances are the biggest considered in the literature so far and the IG-MDA solves them in $329.9$s on average (geo. mean). This data is extracted from the file $\mathtt{results/multiPrim\_grid\_corr\_4d.csv}$ in \citep{code}.
    
    \begin{landscape}
     	\begin{table}
    		\scriptsize{
    			\centering
    			\caption{Instances from \citep{Fernandes2020} with correlated edge costs. Cardinality $|T_G|$ of solution sets, $|\transitionNodes|$ of transition nodes, number of iterations, and time report geometric means built among solved instances only. The speedups consider only instances solved by both algorithms.}
    			\label{tab:corr}
    			\begin{tabular}{R R R R R R R R R R R R R}
    				\multirow{2}{*}{$|V|$} & \multirow{2}{*}{Insts} & \multicolumn{5}{c}{BN} & \multicolumn{5}{c}{IG MDA} & \multirow{2}{*}{Speedup} \\
    				\cmidrule(lr){3-7} \cmidrule(lr){8-12}
    				& & Sol & |T_{G}^*| & |\transitionNodes| & \text{Iterations} & \text{Time} & Sol. & |T_{G}| & |\transitionNodes| & \text{Iterations} & \text{Time} & \\
    				\midrule
    				\multicolumn{13}{c}{COMPLETE $3$d corr. edge costs}\\
    				\midrule
    				%5 & 30 & 30 & 4.21 & 3.95 & 4.65 & 0.0000 & 30 & 4.21 & 3.68 & 7.41 & 0.0000 & 1.45 \\
    				%7 & 30 & 30 & 9.40 & 9.77 & 22.20 & 0.0000 & 30 & 9.40 & 8.44 & 23.80 & 0.0000 & 1.70 \\
    				%10 & 30 & 30 & 34.78 & 54.52 & 332.95 & 0.0004 & 30 & 34.78 & 34.67 & 205.54 & 0.0001 & 3.54 \\
    				%12 & 30 & 30 & 73.27 & 155.52 & 1487.27 & 0.0024 & 30 & 73.27 & 89.00 & 720.29 & 0.0005 & 5.27 \\
    				15 & 30 & 30 & 195.18 & 786.77 & 16508.49 & 0.0438 & 30 & 195.18 & 324.67 & 4988.89 & 0.0038 & 11.50 \\
    				17 & 30 & 30 & 276.97 & 3240.53 & 101568.59 & 0.4329 & 30 & 276.97 & 769.99 & 14026.10 & 0.0149 & 29.09 \\
    				20 & 30 & 22 & 289.40 & 6793.78 & 186354.27 & 0.7750 & 30 & 626.75 & 4364.52 & 109898.99 & 0.2118 & 35.26 \\
    				22 & 30 & 19 & 228.20 & 6070.89 & 154896.66 & 0.5645 & 30 & 847.30 & 10209.11 & 390923.30 & 1.0635 & 24.93 \\
                	\midrule
                	\multicolumn{13}{c}{GRID $3$d corr. edge costs}\\
                	\midrule
                	
%24 & 30 & 30 & 31.99 & 151.65 & 778.06 & 0.0010 & 30 & 31.99 & 55.30 & 333.35 & 0.0002 & 6.14 \\
%27 & 30 & 30 & 39.30 & 272.62 & 1613.77 & 0.0025 & 30 & 39.30 & 99.51 & 677.97 & 0.0004 & 6.29 \\
%30 & 30 & 30 & 46.94 & 438.37 & 2793.53 & 0.0052 & 30 & 46.94 & 122.87 & 969.94 & 0.0005 & 10.33 \\
33 & 30 & 30 & 77.65 & 915.59 & 7991.91 & 0.0185 & 30 & 77.65 & 176.45 & 2042.85 & 0.0013 & 14.42 \\
36 & 30 & 28 & 76.38 & 1521.15 & 15417.74 & 0.0808 & 30 & 93.14 & 342.61 & 4519.75 & 0.0037 & 18.52 \\
38 & 30 & 30 & 56.21 & 396.92 & 3136.16 & 0.0060 & 30 & 56.21 & 89.00 & 966.31 & 0.0005 & 12.56 \\
40 & 30 & 21 & 39.58 & 349.33 & 2406.14 & 0.3498 & 29 & 114.07 & 848.70 & 10929.89 & 0.0180 & 12.46 \\

                \midrule
    			\multicolumn{13}{c}{COMPLETE $4$d corr. edge costs}\\
                \midrule
% 5 & 30 & 30 & 6.45 & 5.12 & 7.07 & 0.0000 & 30 & 6.45 & 4.77 & 11.95 & 0.0000 & 1.50 \\
% 7 & 30 & 30 & 20.16 & 14.05 & 50.21 & 0.0001 & 30 & 20.16 & 11.93 & 56.06 & 0.0000 & 2.02 \\
%10 & 30 & 30 & 102.68 & 99.80 & 1191.57 & 0.0021 & 30 & 102.68 & 67.59 & 738.96 & 0.0006 & 3.51 \\
12 & 30 & 30 & 326.82 & 283.94 & 7713.31 & 0.0319 & 30 & 326.82 & 176.10 & 4076.06 & 0.0057 & 5.64 \\
15 & 30 & 26 & 713.27 & 1646.49 & 81809.87 & 0.7962 & 30 & 1177.38 & 994.76 & 48608.53 & 0.1733 & 14.99 \\
17 & 30 & 20 & 294.13 & 1191.90 & 31441.59 & 0.1965 & 30 & 1448.65 & 2110.19 & 124335.61 & 0.7096 & 11.50 \\
20 & 30 & 13 & 305.19 & 2523.53 & 56573.88 & 0.1986 & 21 & 1444.82 & 3938.87 & 155447.82 & 0.6001 & 25.25 \\
                \midrule
    			\multicolumn{13}{c}{GRID $4$d corr. edge costs}\\
                \midrule
%20 & 30 & 30 & 26.74 & 46.34 & 204.67 & 0.0002 & 30 & 26.74 & 29.46 & 157.61 & 0.0001 & 2.31 \\
24 & 30 & 30 & 89.11 & 690.09 & 5327.06 & 0.0135 & 30 & 89.11 & 221.65 & 2098.39 & 0.0020 & 6.79 \\
27 & 30 & 30 & 167.53 & 1882.84 & 20551.53 & 0.0823 & 30 & 167.53 & 446.02 & 5610.28 & 0.0072 & 11.50 \\
30 & 30 & 28 & 163.55 & 2795.94 & 35350.59 & 0.1349 & 30 & 216.40 & 840.08 & 13791.31 & 0.0216 & 13.05 \\
33 & 30 & 23 & 82.51 & 669.94 & 5933.57 & 0.0141 & 29 & 206.47 & 842.73 & 14760.35 & 0.0222 & 9.27 \\
36 & 30 & 20 & 49.51 & 311.78 & 2518.64 & 0.0057 & 27 & 193.85 & 812.96 & 13934.90 & 0.0225 & 11.20 \\
38 & 30 & 26 & 144.41 & 806.94 & 11284.06 & 0.0347 & 30 & 235.57 & 352.92 & 8144.57 & 0.0112 & 11.46 \\
40 & 30 & 18 & 66.30 & 598.21 & 4758.39 & 0.0153 & 20 & 98.46 & 234.39 & 3118.01 & 0.0036 & 12.27 \\
            \end{tabular}
            }
        \end{table}
    \end{landscape}

    \section{Conclusion}
	The Implicit Graph Multiobjective Dijkstra Algorithm (IG-MDA) is a new dynamic programming algorithm for the Multiobjective Minimum Spanning Tree (MO-MST) problem. For its design we manipulated the transition graph defined in \citet{Santos18} using cost dependent criteria to reduce its size and thus enhance the performance of the used algorithms. Dynamic programming for MO-MST problems entails solving an instance of the One-to-One Multiobjective Shortest Path problem arising transition graph. To solve this instance we use a modified version of the recently published Targeted Multiobjective Dijkstra Algorithm. In this paper, we analyzed the size of the transition graph to motivate its implicit handling in the IG-MDA. Storing the graph explicitly would otherwise lead to unreasonable memory consumption on many instances.
	To benchmark the performance of the IG-MDA, we compared it to our own version of the BN algorithm from \citep{Santos18} on a big set of instances from the literature. The results show that the IG-MDA is more efficient. To the best of our knowledge, it also manages to solve bigger instances than the biggest ones solved so far in the literature.
	
    \bibliographystyle{abbrvnat} 
    \bibliography{literature}
    
    \newpage
    \appendix
    \section{Build Network Algorithm}
    \label{sec:bn}
    In this section, we briefly describe the \emph{Build Network} (BN) algorithm from \citep{Santos18}, using our terminology to easier compare it with the IG-MDA in what follows.

    Again, we start with a given a MO-MST instance $(G,c,s)$ and build the One-to-One MOSP instance $\mathcal{I}_{\text{SP}} = (\transitionGraph, \gamma, \{s\}, V)$. Note that the original transition graph $\transitionGraph$ is used instead of $\transitionGraph^*$. Then, the BN algorithm looks for efficient $\{s\}$-$V$-paths w.r.t. $\gamma$ in $\transitionGraph$. Thereby, for every new explored path $p$ the algorithm analyzes the ordering in which arcs were added to $p$ to determine the allowed expansions of $p$. The conditions ensure that for every spanning tree of $G$ the BN algorithm only considers one $\{s\}$-$V$-path in $\transitionGraph$ that represents it. This is remarkable given that a tree with $k$ nodes can have up to $(k-1)!$ representations in $\transitionGraph$ as noted in \Cref{sec:size}. These unique representations of trees in $\transitionGraph$ are called \emph{minimal paths}.
    
    \begin{definition}[Minimal Paths (cf. Definition 3.5. \citet{Santos18})]
        \label{def:minimalPaths}
        Let $q = (a_1, \dotsc, a_{k})$ be a path in $\transitionGraph$ with $k$ arcs for some $k \leq n-1$. For the arcs $a_i$, we set $a_i^{-1} = [u_i, w_i] \in E$. $q$ is said to be a \emph{minimal path} if it satisfies one of the following conditions:
        \begin{enumerate}
            \item $k=1$
            \item $k \geq 2$, the subpath $p = (a_1, \dotsc, a_{k-1})$ is a minimal path and there holds $u_{k-1} = u_k$, $w_{k-1} < w_{k}$. \label{item:childrenInOrder}
            \item $k \geq 2$, the subpath $p = (a_1, \dotsc, a_{k-1})$ is a minimal path and there exist to indices $\ell, \, \ell' \leq k$ with $\ell < \ell'$ and there holds $u_{k-1} = w_\ell$ and $u_k = w_{\ell'}$.\label{item:parentsInOrder}
        \end{enumerate}
    \end{definition}

    To understand the latter definition we consider the preimage edges $[u_i, w_i]$ of $q$ as ordered pairs. Then, for the leading subpath $p$ of $q$ with $i$ arcs, $i < k$, the node $w_i$ is the $(i+1)$\ts{th} node added to the tree in $G$ represented by $q$. Then, the last condition in \Cref{def:minimalPaths} states, in terms of paths in $\transitionGraph$, that if $w_\ell$ is added to a spanning tree after $w_{\ell'}$, the adjacent nodes of $w_\ell$ must be considered prior to the adjacent nodes of $w_{\ell'}$. Additionally, the second condition, states that for a  node $u$, its adjacent nodes have to be considered in order w.r.t. the considered indexing of the nodes in $G$. Then, the statement in \citet[Proposition 3.7.]{Santos18} is easy to see: there is a one-to-one correspondence between minimal paths in $\transitionGraph$ and trees in $G$.

    The pseudocode of our implementation of the BN algorithm is in \Cref{algo:bn}.

   	\begin{algorithm}
    	\small{
    		\SetKwInOut{Input}{Input} \SetKwInOut{Output}{Output}
    		\Input{$d$-dimensional One-to-One MOSP instance $\mathcal{I}_{\text{SP}} = (\transitionGraph^*, \gamma, \{s\}, V)$.}
    		\BlankLine
    		\Output{Minimal complete set $\mcs{V}$ of efficient $\{s\}$-$V$-paths in $\transitionGraph$ w.r.t. $\gamma$.}
    		\BlankLine
    		Priority queue of paths $Q \leftarrow{} \emptyset{}$\label{algo:bn:initStart}\tcp*{Sorted according to $\gamma$.}
            Transition graph $\transitionGraph$ initialized only containing the transition node $\{s\}$\label{algo:bn:implicitInit}\;
    		Trivial $\{s\}$-$\{s\}$-path $p_{\text{init}} \leftarrow{} ()$\;
    		$Q \leftarrow{} Q.\mathtt{insert}(p_{\text{init}})$\label{algo:bn:insertInitialPath}\;
    		\BlankLine
    		\While{$Q \neq \emptyset$}{
    			$p \leftarrow{} Q.\mathtt{extractMin}()$ \label{algo:bn:extraction}\;
    			$U \leftarrow{}$ last transition node of path $p$ \tcp*[r]{If $p = p_{\text{init}}$, $U \leftarrow \{s\}$.}
                \lIf{$\gamma(\mcs{U}) \dom \gamma(p)$}{\textbf{continue}\label{bn:extractionDomCheck}}
                %$\mathcal{P}_U.\text{append}(p)$\;
                %\If{$\text{exploredTrees}[U] \neq \emptyset$}{
                %$Q.\mathtt{insert}(\arglexmin\{\gamma(p')\ | \ p' \in \text{exploredTrees}[U]\})$\label{algo:bn:newQueuePath}
                %}
    			\BlankLine
                Boolean flag $\mathtt{sucess} \leftarrow $ False\;
                \lIf{$\outgoing{U}$ not initialized}{build $\outgoing{U}$ as described in \Cref{sec:implicitNodesBN} \label{algo:bn:newArcs}}
    			$\mathcal{MIN}_p \leftarrow $ Minimal paths $p \circ a$ for $a \in \outgoing{U}$ \label{algo:bn:minPaths}\;
                \For{$q \in \mathcal{MIN}(p)$} {
                    Assume $q$ is spanning tree of $W \subseteq V$\;
                    \If{\textbf{not }$\gamma(\mathcal{P}_W) \dom \gamma(q)$ \label{algo:bn:dominanceCheck}} {
                        $Q.\mathtt{insert}(q)$\label{algo:bn:insertion}\;
                        $\mathtt{sucess} \leftarrow $ True\;
                    }
                }
                \lIf{$\mathtt{sucess} ==$ True}{$\mathcal{P}_U.\mathtt{append}(p)$}
    		}
    		\Return $\mathcal{P}_{V}$;
    		
    		\caption[caption]{Built Network (BN) algorithm.}\label{algo:bn}}
    \end{algorithm}

    We combine the minimal path definition from \citep{Santos18} with a \emph{lazy queue management approach} for explored paths that notably enhances the running time of label setting MOSP algorithms as experienced in \citep{Ahmadi21, Casas22}. Lazy queue management of explored paths works as follows. 
    Assume the minimal $\{s\}$-$W$-path $q$ in $\transitionGraph$ is considered by the algorithm. 
    It is immediately discarded (\Cref{algo:bn:dominanceCheck}) if it is dominated by or equivalent to any path in $\mcs{W}$.
    Otherwise, it is inserted into the algorithm's priority queue $Q$ without further checks (\Cref{algo:bn:insertion}). 
    This motivates the \emph{lazy} attribute of this method since $q$ is not compared to other explored $\{s\}$-$W$-paths. 
    Instead, because of the lex. ordering of paths in $Q$, $q$ is extracted from $Q$ after every explored $\{s\}$-$W$-path $q'$ with $\gamma(q') \lexsmallereq \gamma(q)$. 
    Note that this also holds for $\{s\}$-$W$-paths that are added later than $q$ but before $q$'s extraction.
    These $q'$ paths are the only explored $\{s\}$-$W$-paths that can dominate $q$ and they are possibly added to $\mcs{W}$ before $q$ is extracted. 
    Thus, our version of the BN algorithm with lazy queue management repeats the dominance or equivalence check $\gamma(\mcs{W}) \dom \gamma(q)$ after $q$'s extraction from $Q$ (\Cref{bn:extractionDomCheck}). 
    Only if $q$ turns out not to be dominated by or equivalent to any path in $\mcs{W}$ after this check, it is further considered by the algorithm. 
    In this case, the BN algorithm needs to evaluate the outgoing arcs in $\outgoing{W}$ s.t. only minimal paths arising from $q$'s expansions are considered (\Cref{algo:bn:minPaths}).

    The correctness of our implementation of the BN algorithm follows from the fact that considering minimal paths in $\transitionGraph$ suffices to find a minimum complete set of efficient paths that represent a minimum complete set of efficient spanning trees in $G$ \citep[Proposition 3.7.]{Santos18} and from the correctness of the NAMO$A^*$-lazy algorithm introduced in \citep{Casas22}. All in all, we have designed a new algorithm combining the elegant pruning rule (minimal paths) that made the original BN algorithm state of the art with recent advances used to improve the handling of explored paths in label setting MOSP algorithms.

    \subsection{Implicit Handling of the Transition Graph}
    \label{sec:implicitNodesBN}

    Handling the transition graph $\transitionGraph$ implicitly is easier for the BN algorithm than for the IG-MDA. It works as described in \Cref{sec:implicitNodes} with the following two differences. First, neither lists of incoming arcs nor $\NQP$ lists have to be maintained. Second, the lists of $\outgoing{U}$ of outgoing arcs for a transition node $U$ are built in \Cref{algo:bn:newArcs} as described in \eqref{eq:outgoingArcs} without any arc removal conditions. For every $\{s\}$-$U$-path $p$ extracted from $Q$ the relevant outgoing arcs needed to build minimal path expansions of $p$ are chosen in a path-dependent way in \Cref{algo:bn:minPaths}. We suppose that the authors of \citet{Santos18} also generate the transition graph $\transitionGraph$ implicitly as described in this section to restrict the memory consumption of $\transitionGraph$ as far as possible.

    \subsection{Comparison to the IG MDA}
 
    There are two main differences between the IG-MDA and our new BN algorithm: 
    \begin{description}
        \item[Explored paths] The IG-MDA algorithm restricts the size of the queue and stores other explored paths in $\NQP$ lists. It requires to search for a new queue path in every iteration but does not require a dominance or equivalence check after a path's extraction from the queue. Our new BN algorithm stores multiple explored paths per node in the queue but requires dominance or equivalence checks right after every path's extraction from the queue.
        \item[Pruning conditions] The IG-MDA uses cost dependent criteria (\Cref{lem:efficientParallelArcs}, \Cref{lem:efficientCutExit}) to reduce the number of arcs in the transition graph in a path-independent way. The BN algorithm in its original version as well as in our implementation uses path-dependent pruning techniques (minimal paths, cf. \Cref{def:minimalPaths}) forcing the inclusion of nodes in the represented spanning trees to follow a fixed order/rule.
    \end{description}

    \begin{example}[Different Pruning Criteria]
        The path $p = ((1,2), (1,3), (1,4))$ is minimal according to \Cref{def:minimalPaths} and represents the same tree than the path $p' = ((1,2), (1,4), (1,3))$. The BN algorithm would discard the expansion of the subpath $((1,2), (1,4))$ along the arc $(1,3)$ because it violates \Cref{item:childrenInOrder} in \Cref{def:minimalPaths}. The IG-MDA allows this expansion of the path if the arc $(1,3)$ is not dominated by any arc in the cut defined by the node set $\{1, 2, 4\}$. Note however that at the transition node $W = \{1, 2, 3, 4\}$ $p$ and $p'$ meet with equivalent costs and the $\dom$ operator guarantees that only one is kept and further expanded.
    \end{example}

    \subsection{Efficienciy of our BN implementation}
    \label{sec:bnEfficiency}
    We tried to determine whether our implementation of the BN algorithm from \citep{Santos18} is efficient. Since we do not have access to the original implementation from the authors, we extrapolated their computational results to the performance on our machine. Using \citep{CPUBenchmark} we found out that on single-threaded jobs the processor used in \citep{Santos18} delivers $76.3\%$ of the performance of our processor. Thus we scaled the average running times for the SPACYC based instances in Table 6 of \citep{Santos18} accordingly to obtain the expected running times of their code on our machine. In \Cref{tab:runningTimesCPUcorrected} we list the original running times from \citep{Santos18}, the CPU-scaled running times we obtained, and the running times we obtained from our implementation of \Cref{algo:bn}.     Since the BN implementation in \citep{Santos18} is coded in Java and ours is coded in C++, the running times are still not completely comparable. However, our running times seem to be good enough ($\times6$ faster on the biggest $3$d instances and $\times1.6$ faster on the biggest $4$d instances) to claim that our implementation of the BN algorithm is efficient and useful for the comparisons that follow in this paper. The speedup decreases notably with increasing edge cost dimension because the impact of dominance checks on running time increases in higher dimensions. Most probably, both programming languages are similarly efficient in performing these checks. Note that the reported running time averages in \Cref{tab:runningTimesCPUcorrected} do not coincide with those reported in \Cref{tab:SPACYC}. The reason is that in \citep{Santos18} the authors always use arithmetic means and in this paper we use geometric means everywhere except in \Cref{tab:runningTimesCPUcorrected}.
    
    \begin{table}
        \small{
        \centering{
        \caption{Comparison of BN algorithm implementations in \citep{Santos18} and in \citep{code}. Computations using $3$ and $4$ dimensional SPACYC based instances.}        \label{tab:runningTimesCPUcorrected}
        \begin{tabular}{c | R R R R R}
            Nodes & 10 & 11 & 12 & 13 & 14 \\
            \midrule
            BN $3$d in \citep{Santos18} & 0.25 & 1.00 & 5.42 & 22.11 & 86.33\\
            CPU scaled & 0.19 & 0.76 & 4.14 & 16.87 & 65.87\\
            BN $3$d as in \Cref{algo:bn} & 0.04 & 0.10 & 0.54 & 2.19 & 11.25
            \\
            \midrule
            BN $4$d in \citep{Santos18} & 5.58 & 28.40 & 188.14 & 712.96 & 3656.91\\
            CPU scaled & 4.26 & 21.67 & 143.55 & 543.99 & 2790.22\\
            BN $4$d as in \Cref{algo:bn} & 0.76 & 6.34 & 69.76 & 351.73 & 1737.88
        \end{tabular}
        }
        }
    \end{table}
 
\end{document}